\documentclass[fleqn,usenatbib]{mnras}

\usepackage{newtxtext,newtxmath}
\usepackage{pdflscape}

\usepackage[T1]{fontenc}
\usepackage{ae,aecompl}

\usepackage{graphicx}	
\usepackage{amsmath}	
\usepackage{amssymb}	

\title[Hosts of MWA compact sources]{Interplanetary Scintillation studies with the Murchison Wide-field Array IV: The hosts of sub-arcsecond compact sources at low radio frequencies}

\author[Elaine M. Sadler et al.]{
Elaine M. Sadler,$^{1,2,3}$\thanks{E-mail:elaine.sadler@sydney.edu.au (EMS)}
Rajan Chhetri,$^{2,4}$
John Morgan,$^{4}$
Elizabeth K. Mahony,$^{1,2,3}$
\newauthor
Thomas H.\ Jarrett$^{5}$ and Steven Tingay$^{4}$\\
$^{1}$Sydney Institute for Astronomy, School of Physics A28, The University of Sydney, NSW 2006, Australia \\
$^{2}$ARC Centre of Excellence for All-Sky Astrophysics (CAASTRO) \\
$^{3}$CSIRO Astronomy and Space Science, PO Box 76, Epping, NSW 1710, Australia\\
$^{4}$International Centre for Radio Astronomy Research, Curtin University, GPO Box U1987, Perth, WA 6845, Australia\\
$^{5}$Department of Astronomy, University of Cape Town, Private Bag X3, Rondebosch, 7701, South Africa
}

\date{Accepted XXX. Received YYY; in original form ZZZ}

\pubyear{2018}

\begin{document}
\label{firstpage}
\pagerange{\pageref{firstpage}--\pageref{lastpage}}
\maketitle

\begin{abstract}
Around 10\% of bright low-frequency radio sources observed with the Murchison Widefield Array (MWA) show strong Interplanetary Scintillation (IPS) on timescales of a few seconds, implying that almost all their low-frequency radio emission comes from a compact component less than 0.5\,arcsec in angular size. Most of these objects are compact steep-spectrum (CSS) or MHz-peaked spectrum (MPS) radio sources. 
We have used mid-infrared data from the Wide-field Infrared Survey Explorer (WISE) catalogue to search for the host galaxies of 65 strongly-scintillating MWA sources and compare their properties with those of the overall population of bright low-frequency radio sources. 
We identified WISE mid-infrared counterparts for 91\% of the bright sources in a single 900 deg$^2$ MWA field, and found that the hosts of the strongly-scintillating sources were typically at least 1\,mag fainter in the WISE W1 (3.4\,$\mu$m) band than the hosts of weakly-scintillating MWA sources of similar radio flux density. This difference arises mainly because the strongly-scintillating sources are more distant. We estimate that strongly-scintillating MWA sources have a median redshift of $z\sim1.5$, and that at least 30\% of them are likely to lie at $z>2$. The recently-developed wide-field IPS technique therefore has the potential to provide a powerful new tool for identifying high-redshift radio galaxies without the need for radio spectral-index selection. 
\end{abstract}




\section{Introduction}

\cite{Morgan2018} recently developed a novel technique that uses high time-resolution imaging with the Murchison Widefield Array (MWA; \cite{Tingay2013}) to measure interplanetary scintillation (IPS) for many hundreds of  radio sources simultaneously across a wide area of sky. Since strong IPS is only seen for sources that are dominated by a compact component less than $\sim0.5$\,arcsec in size, this technique makes it possible to identify large numbers of sub-arcsecond radio sources efficiently without the need for VLBI observations.

\cite{Chhetri2018} applied this technique to a five-minute observation of a 900 deg$^2$ MWA field at 162\,MHz with 0.5\,s time resolution, and found that 12\% of the strong continuum sources in this field showed rapid fluctuations caused by IPS. Interestingly, \cite{Chhetri2018} showed that the compact source population at low frequencies is dominated by peaked-spectrum radio sources, and that many of the flat-spectrum QSOs that are compact at frequencies of a few GHz show extended structure at 160\,MHz. All the sources studied by \cite{Chhetri2018} are also catalogued in the 72--231\,MHz MWA GLEAM survey \citep{Wayth2015, Hurley-Walker2017}. 

Our aim in this paper is to identify the host galaxies of the compact low-frequency sources found by \cite{Chhetri2018}, and to compare their properties with those of the host galaxies of other bright MWA sources. Since the \cite{Chhetri2018} sample contains a significant number of peaked-spectrum radio sources, we can also test the recent suggestion \citep{Coppejans2015,Callingham2017} that many MHz peaked-spectrum (MPS) radio sources are high-redshift objects at $z>2$. 

We take as a reference point the detailed spectroscopic studies of low-frequency radio sources listed in Table\,\ref{tab:review}. Together, these provide a broad picture of the typical optical hosts (and redshift distribution) of bright (flux density above $\sim1$\,Jy), low-frequency radio sources. 

At flux densities above 5--10\,Jy, the 178\,MHz-selected sample of \cite{Laing1983} and the 408\,MHz-selected sample of \cite{Best1999} each provide samples of $\sim200$ well-studied sources. 
Both studies have an optical identification rate close to 100\% and near-complete optical spectroscopic information on the radio-source hosts. Despite the difference in selection frequency, both samples contain a similar mix of 70--75\% radio galaxies and 25-30\% radio-loud QSOs and have a very similar median redshift ($z\sim0.48$). It therefore seems reasonable to assume that the overall properties of radio sources selected at $\sim150$\,MHz should be broadly similar to those of samples selected at 408\,MHz. 

The 408\,MHz MRC-1\,Jy survey \citep{McCarthy1996,Kapahi1998b,Baker1999} probes a flux-density range similar to that of the \cite{Chhetri2018} 162\,MHz sample. As can be seen from Table \ref{tab:review}, 96\% of the sources in the MRC-1\,Jy sample have been optically identified. 
Roughly 80\% of the MRC-1\,Jy sources are radio galaxies, and 20\% are radio-loud QSOs. 

\begin{table*}
  \caption{Summary of large-area spectroscopic samples of bright low-frequency radio sources. The location of each sample on the sky area is indicated in column 3 as: N = Northern hemisphere, S = Southern hemisphere, E = Equatorial. Our IPS field overlaps with parts of the MRC-BRL, MS4 and MRC-1Jy survey areas.  }
  \label{tab:review}
\begin{tabular}{lccrrllll} 
\hline
Sample & Freq. & Sky & Flux density & Sources & References & Notes & Median \\
       & (MHz) & area & limit (Jy) & & & & redshift \\
\hline
 3CR &  178 & N & 10.9 & 205 & \cite{Laing1983} & Optical ID rate 96\% & 0.48 \\
     &      & & & & & (71\% galaxies, 25\% QSOs) \\
   &&&&& \\
MRC-BRL & 408 & E & 5.0 & 178 & \cite{Best1999} & Optical ID rate 100\% & 0.47 \\
 & & & & & & Spectroscopic completeness 98\% \\
& & & & & & (72\% galaxies, 28\% QSOs) & \\
 &&&&& \\
MS4  & 408  & S & 4.0 & 228 & 
\cite{Burgess2006a,Burgess2006b} & Optical ID rate $>90\%$  & $\sim$0.5 \\
 &&&&& \\
MRC-1\,Jy & 408 & S & 0.95 & 558 & \cite{McCarthy1996};  & Optical ID rate 96\% & 0.63 \\ 
& & & & & \cite{Kapahi1998b} and  & (80\% galaxies, 20\% QSOs) &  \\
& & & & & \cite{Baker1999}\\
 &&&&& \\
\hline
\end{tabular}
\end{table*}

\section{Sample selection and WISE cross-matching} 
\subsection{Overview}
The data samples analysed in this paper are drawn from the 900\,deg$^2$ MWA field observed by \cite{Chhetri2018}.
To identify the host galaxies of these sources, we used the all-sky mid-IR Wide-field Infrared Survey Explorer (WISE) allWISE catalogue \citep{Wright2010,Cutri2013}. This is currently the deepest large-area photometric catalogue available, and most powerful radio AGN out to redshift $z\sim1.5$ are expected to have a counterpart in the WISE catalogue \citep{Gurkan2014, Glowacki2018}. 

The relatively high surface density of WISE sources (up to 10,000 deg$^{-2}$ at high Galactic latitude) means that sub-arcsec radio positions are needed for reliable cross-matching. Achieving this can be challenging for extended low-frequency sources, which often have complex radio structures with no prominent central core. We therefore took a two-step approach to cross-matching the \cite{Chhetri2018} MWA sources. 

\begin{figure}
\includegraphics[width=\columnwidth]{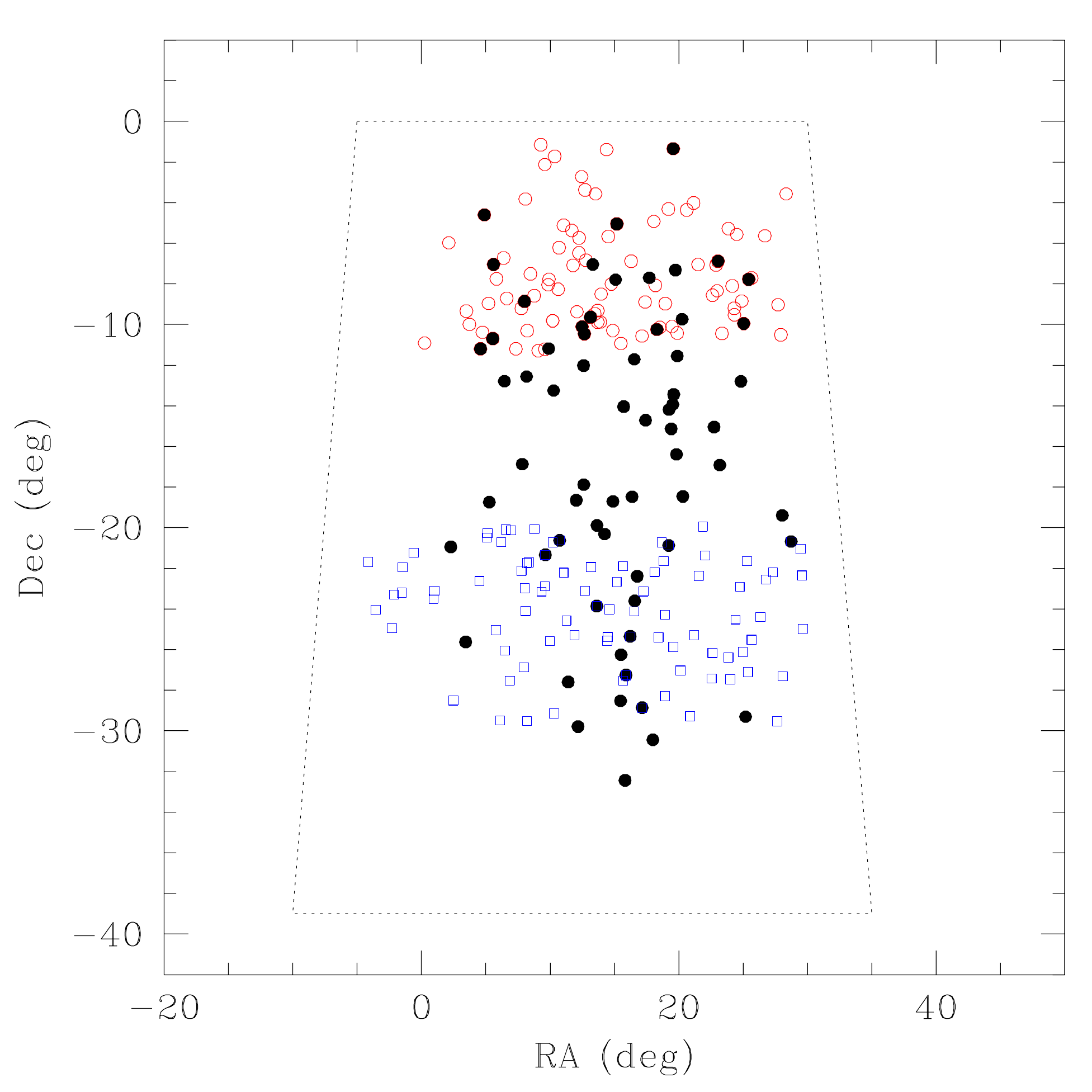}
  \caption{Distribution on the sky of the three data samples defined in \S2.1. Sources in the {\it main IPS-FIRST}\ sample are shown by red open circles, those in the {\it expanded compact}\ sample by black filled circles, and those in the {\it IPS-MRC-1Jy}\ sample by blue open squares. Dotted lines mark out the full area of the MWA IPS field. }
    \label{fig:ips_coords}
\end{figure}

\begin{enumerate}
\item
We first compiled a {\it main IPS-FIRST sample}\ of MWA sources from Table 1 of \cite{Chhetri2018}, selected to have high enough S/N that they can be reliably split into scintillation classes. This sample includes many sources with complex or extended radio structure, so we used the VLA FIRST survey as the basis for our WISE cross-matching as described in \S2.2 below. Since the FIRST survey overlaps only the northern part of the MWA field, the main sample contains only the 88 high S/N sources that lie in this overlap area (see Figure \ref{fig:ips_coords}). 
\item 
We also compiled an {\it expanded compact}\ sample of sources that show strong IPS as defined by \cite{Chhetri2018}.  These strongly-scintillating sources are known to be compact on sub-arcsec scales at 162\,MHz, and experience shows that identifications can be made fairly reliably from the lower-resolution NVSS data. For these compact objects therefore, we can relax both the SNR limit and the restriction that sources need to lie within the FIRST overlap area. The expanded compact sample comprises 65 strongly-scintillating sources, as discussed in \S2.3 below.  
\item
In addition, there are 85 \cite{Chhetri2018} high S/N (normSNR$\geq$12.5) sources in the MRC-1Jy catalogue \citep{McCarthy1996,Kapahi1998b}, which covers two strips of sky between -20$^\circ$ and -30$^\circ$ declination. This {\it IPS-MRC-1Jy sample} has a different selection frequency from the other two samples (408\,MHz rather than 162\,MHz), and as a result it is skewed towards slightly brighter sources than the other two sub-samples. It does however contain additional information that will be valuable for later analysis, including accurate radio positions, galaxy/QSO classifications for all sources, and over 40 sources with published spectroscopic redshifts. 
\end{enumerate}

The overall properties of these three samples are summarised in Table \ref{tab:sample_overview} (where columns 6 and 7 show the number of sources classified as either compact or peaked-spectrum by \cite{Chhetri2018} and column 8 the number of sources detected in the Australia Telescope 20 GHz (AT20G) survey \citep{Murphy2010}). Figure \ref{fig:ips_coords} shows the sky distribution of the three samples described above. 

\begin{table*}
	\centering
	\caption{Overview of data samples used. All sources are selected from Table 1 of Chhetri et al. (2018a), and compact sources are defined to be those with a normalised scintillation index (NSI) $\geq0.9$, meaning that 90\% of the low-frequency flux density arises from a sub-arcsec compact component. The lower S/N cutoff adopted for the expanded compact sample means that this sample contains more faint sources and so has a lower median flux density.}  
	\label{tab:sample_overview}
\begin{tabular}{lrlrccccll} 
\hline
Sample & normSNR & Area & NSI & n & Compact & Peak & AT20G & Median & Notes \\
       &         &      &     &   &    &   &    & $\alpha_{162}^{1400}$ & \\
\hline
Main IPS-FIRST  & $\geq12.5$ & FIRST overlap & All & 88 & 14 & 8  & 20 & -0.84 & S$_{\rm 162} \geq 1.08$\,Jy  \\
sample  & & area only & & & (16\%) & (9\%) & (23\%) &  & (median = 2.28\,Jy)  \\
&&&&& \\
Expanded compact & $\geq$8.0 & Full IPS field & $\geq0.90$ & 65 & 65 & 27 & 15 & -0.67 & S$_{\rm 162} \geq 0.57$\,Jy  \\
sample  & & & & & (100\%) & (42\%) & (23\%) &  & (median = 1.34\,Jy)  \\
&&&&& \\
IPS-MRC-1\,Jy  & $\geq$12.5 & MRC-1\,Jy & All & 85 & 10 &  7 & 18 & -0.78 & S$_{\rm 162} \geq 1.05$\,Jy  \\
sample  & & overlap area & & & (12\%) & (8\%) & (21\%) &  & (median = 2.64\,Jy)  \\
		\hline
	\end{tabular}
\end{table*}

\subsection{Radio cross-matching}
To derive the most accurate radio positions for WISE cross-matching, we matched the MWA GLEAM sources with objects in the higher-resolution 1.4\,GHz FIRST \citep{Becker1995} and NVSS \citep{Condon1998} catalogues. This cross-matching process is reasonably straightforward, since the MWA sources are bright enough at 162\,MHz that they should be easily detectable in the higher-frequency surveys even if resolved into two or more components.\footnote{As noted by \cite{Chhetri2018}, the one exception in their sample is the steep-spectrum cluster relic GLEAM J004130-092221 (MRC 0038-096), which is resolved out and undetected in the FIRST survey. }

For the {\it main IPS-FIRST sample}, we examined the FIRST catalogue and images for all the `high S/N' (normSNR $\geq12.5$) sources from Table 1 of \cite{Chhetri2018} that lay within the FIRST overlap area. Where two or more FIRST sources lay within the MWA beam, we used well-established criteria (see e.g. section 3.3 of \cite{Ching2017}) to associate double and triple sources and measure a radio centroid.  
Of the 88 sources in Table \ref{tab:data2}, 37 have a single FIRST component, 32 are resolved doubles in FIRST and 19 are associated with three or more FIRST components. For sources matched with a single FIRST component, the median offset between the FIRST and MWA GLEAM positions was 3.2\,arcsec. 

For the {\it expanded compact sample}, where all the MWA sources are expected to be single and unresolved on arcsec scales, we carried out a simple cross-match with the NVSS catalogue. All the objects in this sample were reliably matched with a single NVSS source, and for these bright sources the NVSS positional accuracy is typically better than 1.5\,arcsec. The median offset between the NVSS and MWA GLEAM positions was 3.4\,arcsec. 

For the {\it IPS-MRC-1Jy sample}, we used the radio core positions measured by \cite{McCarthy1996} and \cite{Kapahi1998b}, which are typically accurate to better than 1\,arcsec. 

\subsection{The main IPS-FIRST sample}
We restricted our main sample to the 88 high S/N sources from \cite{Chhetri2018} that lie within the 1.4 GHz VLA FIRST survey area. This allows us to use the high-resolution FIRST data to characterise the radio structure, and to measure a more accurate radio centroid for sources with extended radio structure. 
These radio sources are bright enough that most of them can be reliably divided into three IPS classes on the basis of their normalised scintillation index (NSI), as discussed in \S4.1 of \cite{Chhetri2018} (a few high S/N sources have NSI upper limits between 0.4 and 0.6, and so may be either weak or moderate scintillators).

\begin{itemize}
\item
10\% are strong scintillators with NSI $\geq0.9$, in which all the observed flux density arises from a region less than 0.5\,arcsec in angular size, 
\item
27\% are moderate scintillators, with $0.4 \leq {\rm NSI} <0.9$, where most of the observed flux density arises from one or more sub-arcsec compact components, 
and 
\item
63\% show weak or no scintillation, with NSI $<0.4$, and most of the observed flux density is extended on scales of a few arcsec or larger. 
\end{itemize}

\begin{figure*}
\includegraphics[width=15.0cm]{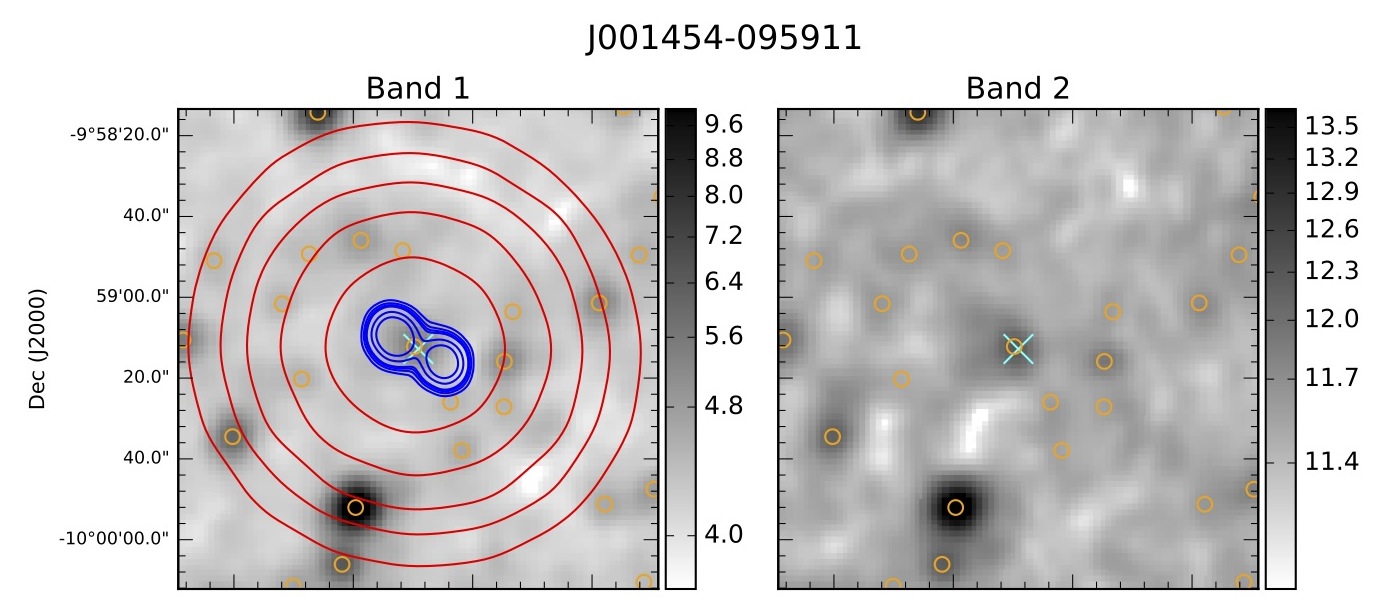}
\includegraphics[width=15.0cm]{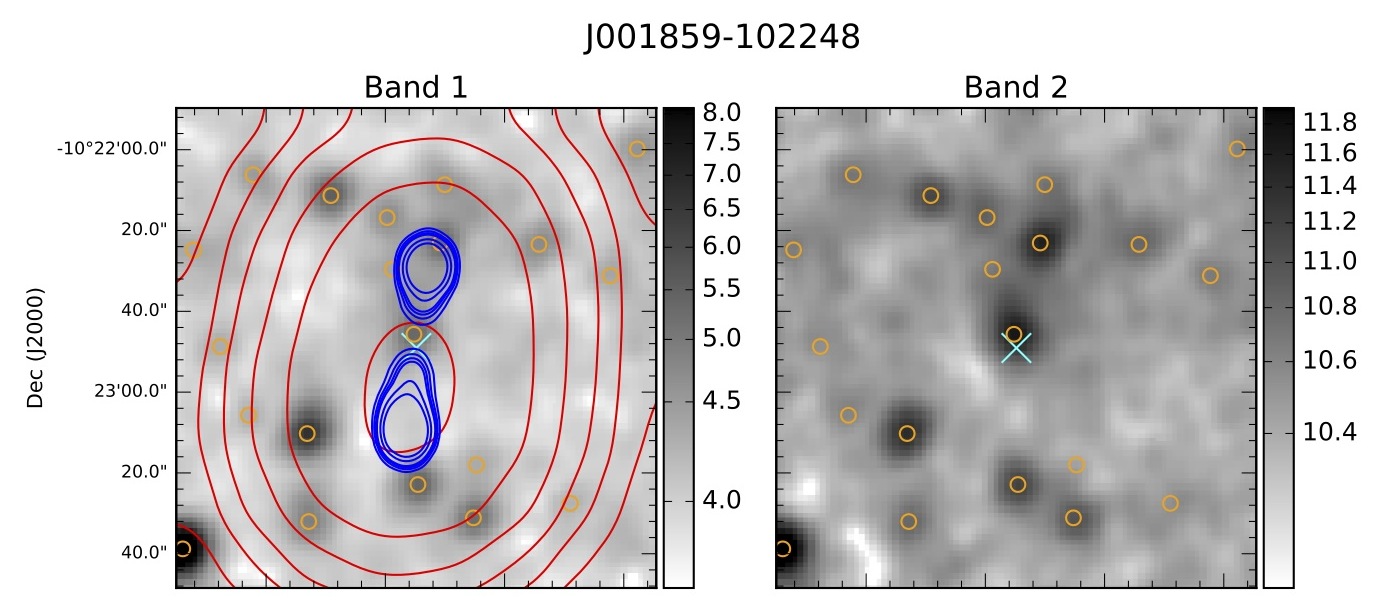}
  \caption{Examples of the overlay plots used to assist  identification of MWA sources. Greyscale images show the WISE W1 (3.4\,$\mu$m) and W2 (4.6\,$\mu$m) data, with radio contours overlaid on the W1 image. Red contours correspond to the 162\,MHz MWA data and blue contours to the higher-resolution VLA FIRST data. A cyan cross marks the FIRST radio centroid, and orange circles mark mid-IR sources listed in the AllWISE catalogue \citep{Cutri2013}. For both MWA sources shown here, the WISE source closest to the radio centroid was accepted as a genuine ID. 
}
    \label{fig:example_overlays}
\end{figure*}

Well-developed and reliable techniques already exist for matching FIRST radio data with large optical surveys \citep[e.g.][]{Best2005,Ching2017}. In this paper we use the same methodology as \cite{Ching2017}, including their technique for associating components of multi-component sources. 

For sources with a single FIRST component (42\% of the main sample), we used a 2.5\,arcsec matching radius to search for a WISE counterpart. For sources with multiple FIRST components, we used the FIRST radio centroid with the same 2.5\,arcsec matching radius, and also made a visual inspection of overlay plots like those shown in Figure \ref{fig:example_overlays}. 

We found WISE matches for 80/88 (91\%) of sources in the main IPS-FIRST sample. 
Notes on some individual sources are in the Appendix. 

\subsection{The expanded compact source sample}
An expanded sample of strongly-scintillating sources, also from Table 1 of \cite{Chhetri2018}, was selected to have normSNR $\geq$8.0 and NSI $\geq$0.90.  We were able to relax the S/N cut for these compact sources, since the high level of scintillation they show can be reliably detected even for weaker sources. The final `expanded compact' sample contains 65 objects, of which 18 are also in the main IPS-FIRST sample and a further eight are in the IPS-MRC-1Jy sample.

Not all the sources in the expanded compact sample lie in the FIRST overlap area, but the compact nature of these sources means that we can use the radio positions from the lower-resolution NVSS survey to match with WISE (again using a 2.5 arcsec matching radius). 

This initial comparison found WISE matches for 53/65 (82\%) of sources in the main sample. For objects with no catalogued WISE source within 2.5\,arcsec of the radio position, one of us (THJ) inspected WISE images of the region around the radio-source position to: (i) search for any fainter, uncatalogued WISE sources, and (ii) check for any objects with AGN-like WISE colours (W1-W2 $>0.60$\, mag) with offsets slightly larger than 2.5\,arcsec. This process identified six further WISE counterparts, three of which were significantly fainter in W1 than the allWISE catalogue limit, giving a final WISE identification rate of 59/65 (91\%).  

\subsection{The IPS-MRC-1Jy sample}
The \cite{Chhetri2018} IPS field overlaps part of the MRC-1\,Jy catalogue area at $-20^\circ$ to $-30^\circ$ declination. There are 85 high S/N (normSNR$\geq$12.5) sources from the \cite{Chhetri2018} catalogue that are also in the MRC-1Jy radio galaxy \citep{McCarthy1996} or QSO \citep{Kapahi1998b} catalogue, and the accurate radio positions available for the MRC-1Jy sources \citep{Kapahi1998a} allow us to make reliable cross-matches with the WISE catalogue. We found WISE matches for 81/85 (95\%) of the MRC-1Jy subsample.

\subsection{Data tables }
Tables \ref{tab:data2}, \ref{tab:data3} and \ref{tab:data4} provide information for the main IPS-FIRST sample, the expanded compact source sample and the IPS-MRC-1\,Jy sample respectively. \\
 
\noindent 
The redshift references in Tables \ref{tab:data2} (column 15), \ref{tab:data3} (column 13) and \ref{tab:data4} (column 14) are coded as follows: 
\begin{tabbing}
6dFGS: ~\= 6dF Galaxy Survey \citep{Jones2009} \\
Ba99: \> \cite{Baker1999} \\
Du89: \>  \cite{Dunlop1989} \\
Dr03: \> \cite{Drake2003} \\
Ho03: \> \cite{Hook2003} \\
Ka98: \> \cite{Kapahi1998b} \\
Mc91: \>  \cite{McCarthy1991} \\
Mc96: \> \cite{McCarthy1996} \\
Os94: \> \cite{Osmer1994} \\
RC3: \> Third Reference Catalogue of Bright Galaxies \\     \> \citep{RC3} \\
Sc65: \> \cite{Schmidt1965} \\
SDSS: \> Sloan Digital Sky Survey DR14 \\
 \> \citep{Abolfathi2018} \\
Ti11: \> \cite{Titov2011} \\
We99: \> \cite{Wegner1999} \\
Wi76: \> \cite{Wills1976} \\
Wo00: \> \cite{Wold2000} \\
Wr83: \> \cite{Wright1983} \\
\end{tabbing}

\section{Results}

\subsection{Mid-IR properties of sources in the IPS field}

\subsubsection{Overall mid-IR properties}
\begin{figure}
	\includegraphics[width=\columnwidth]{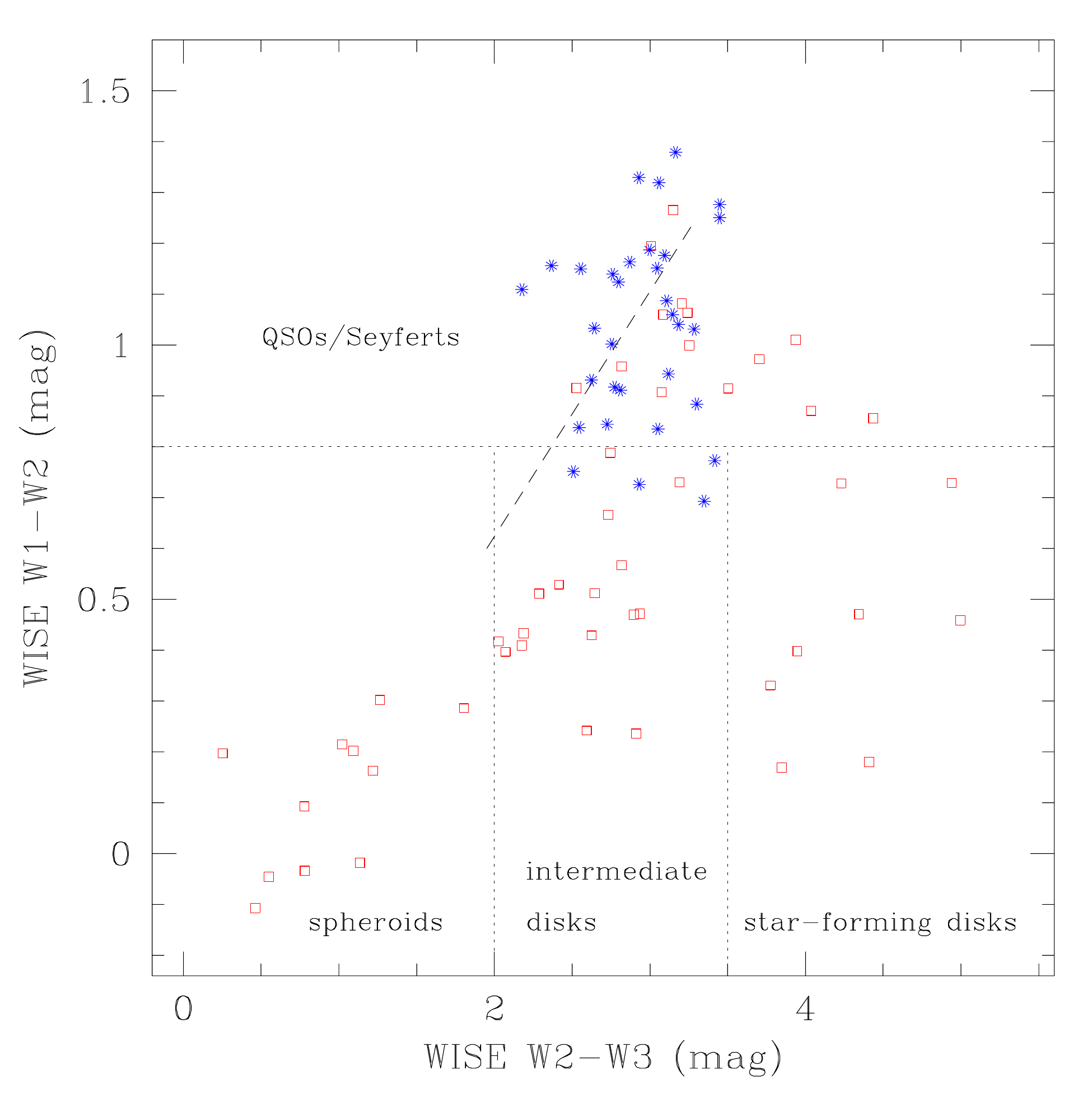}
  \caption{A WISE two-colour diagram \citep{Wright2010} for objects in Tables \ref{tab:data2}, \ref{tab:data3} and \ref{tab:data4} that have reliable detections in all three WISE bands W1 (3.4$\mu$m), W2 (4.6$\mu$m) and W3 (12$\mu$m). Blue points show QSOs, and galaxies are shown by open red squares. The diagonal dashed line marks the `WISE blazar strip' occupied by blazars and radio-loud QSOs \citep{Massaro2012}.  }
    \label{fig:wise_2colx}
\end{figure}

\begin{figure}
	\includegraphics[width=\columnwidth]{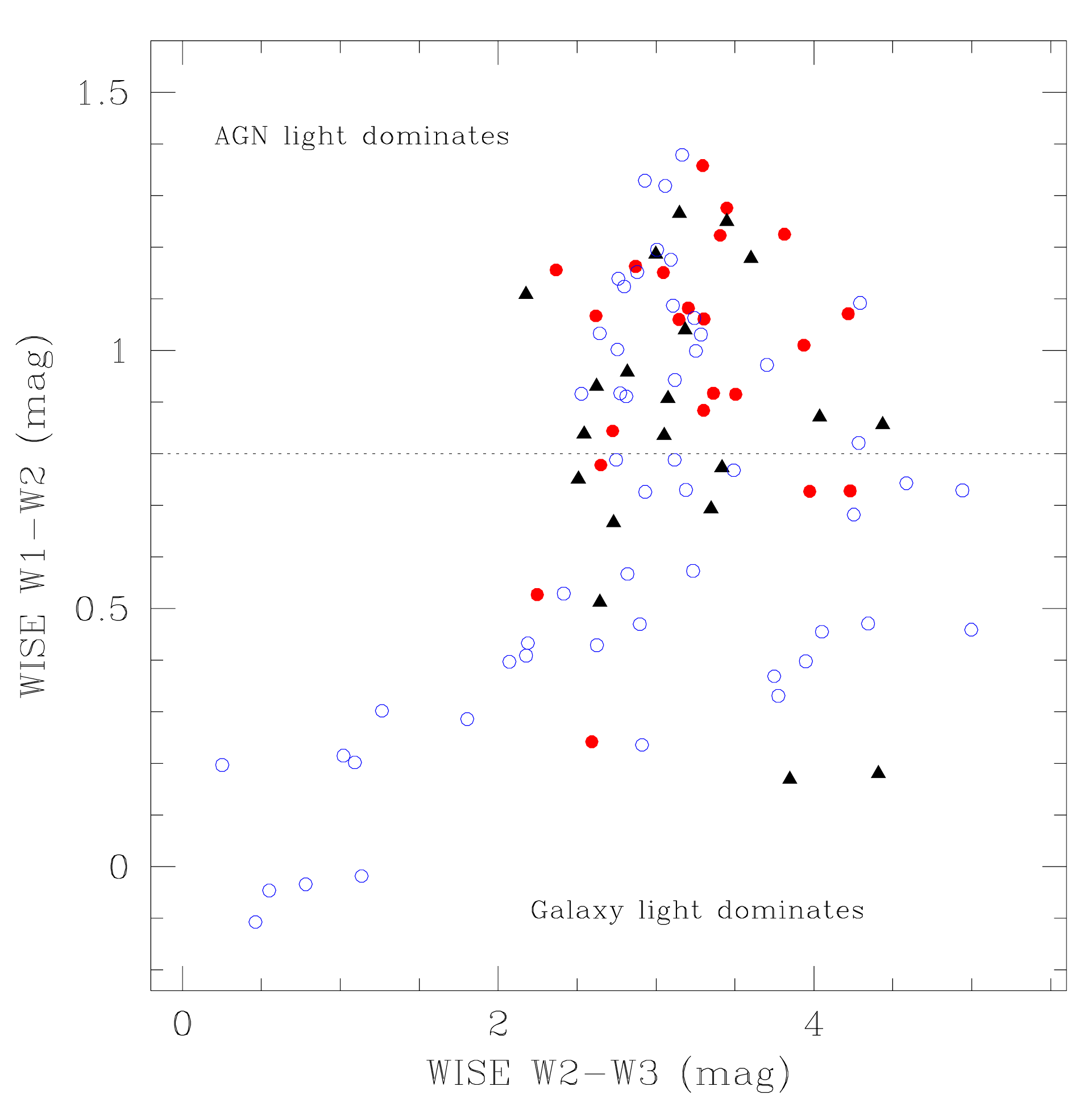}
  \caption{A WISE two-colour diagram similar to Figure \ref{fig:wise_2colx}, but now split into strongly-scintillating sources (red filled circles), moderately scintillating sources (black triangles), and sources showing weak or no scintillation (open blue circles).     }
    \label{fig:wise_2col_scint}
\end{figure}

The WISE two-colour diagrams shown in Figures
\ref{fig:wise_2colx} and \ref{fig:wise_2col_scint} provide a first look at the mid-IR properties of bright low-frequency radio sources out to redshift $z\sim1$ and beyond. 

\cite{Jarrett2017} have shown that WISE colors can provide useful insights into extragalactic radio sources and their evolutionary state. Normal galaxies where the mid-IR emission is dominated by starlight typically have W1--W2 $<0.6$ mag \citep{Wright2010}, with early-type galaxies generally having low W2-W3 colours (W2--W3 $<1.5$) and star-forming galaxies higher values (W2--W3 $>1.0$). Very dusty star-forming galaxies and ULIRGs typically have W2--W3 $>4$. QSOs and other powerful `radiative-mode' AGN have W1--W2 $>0.8$, implying that light from the AGN dominates the mid-IR SED. 

As can be seen from Figure \ref{fig:wise_2colx},  the radio-loud QSOs detected at 162\,MHz span a fairly narrow range in WISE two-colour space, and mainly lie near the WISE `blazar strip' defined by \cite{Massaro2012}. In contrast, the radio galaxies show a large spread across the two-colour diagram, and include both objects with colours typical of galaxies and objects with QSO-like mid-IR colours. 

Figure \ref{fig:wise_2col_scint} shows a WISE two-colour plot split by scintillation class. Most of the strongly-scintillating and moderately-scintillating sources lie in the region with W1--W2 $>0.8$ mag where the AGN light outshines the host galaxy, implying that most of these objects are radiatively-efficient AGN (high-excitation radio galaxies and QSOs; see e.g. \cite{Heckman2014}). 
In contrast, over half of the more extended radio sources with weak or no scintillation have W1--W2 $<0.8$ mag, implying that they are radiatively inefficient (low-excitation, or`jet-mode') AGN whose mid-IR emission is dominated by stellar light from the host galaxy \citep{Ching2017}.  

At this stage, the main conclusion that can be drawn from Figures \ref{fig:wise_2colx} and \ref{fig:wise_2col_scint}
is that most of the strongly-scintillating IPS sources appear to be high-excitation radio galaxies (HERGs; \cite{Heckman2014}) or radio-loud QSOs. 

\subsubsection{The main IPS-FIRST sample} 
Of the 88 main-sample sources in the FIRST overlap area (Table \ref{tab:data2}), 80 (91\%) are reliably matched with a catalogued WISE source but only 29 (33\%) have a known redshift. The mid-IR properties of these radio sources should be typical of the overall MWA source population at flux densities near $\sim1$\,Jy. 

Of the 80 sources with WISE IDs, 37 (46\%) have W1--W2 colours $>$0.8, i.e. consistent with a radiatively-efficient AGN, and 43 (54\%) have W1--W2 colours $\leq$0.8, i.e. dominated by galaxy stellar light. Thus the main sample contains a roughly equal mixture of `radiative mode' and `jet mode' radio AGN \citep{Heckman2014}. 

\begin{figure}
	\includegraphics[width=\columnwidth]{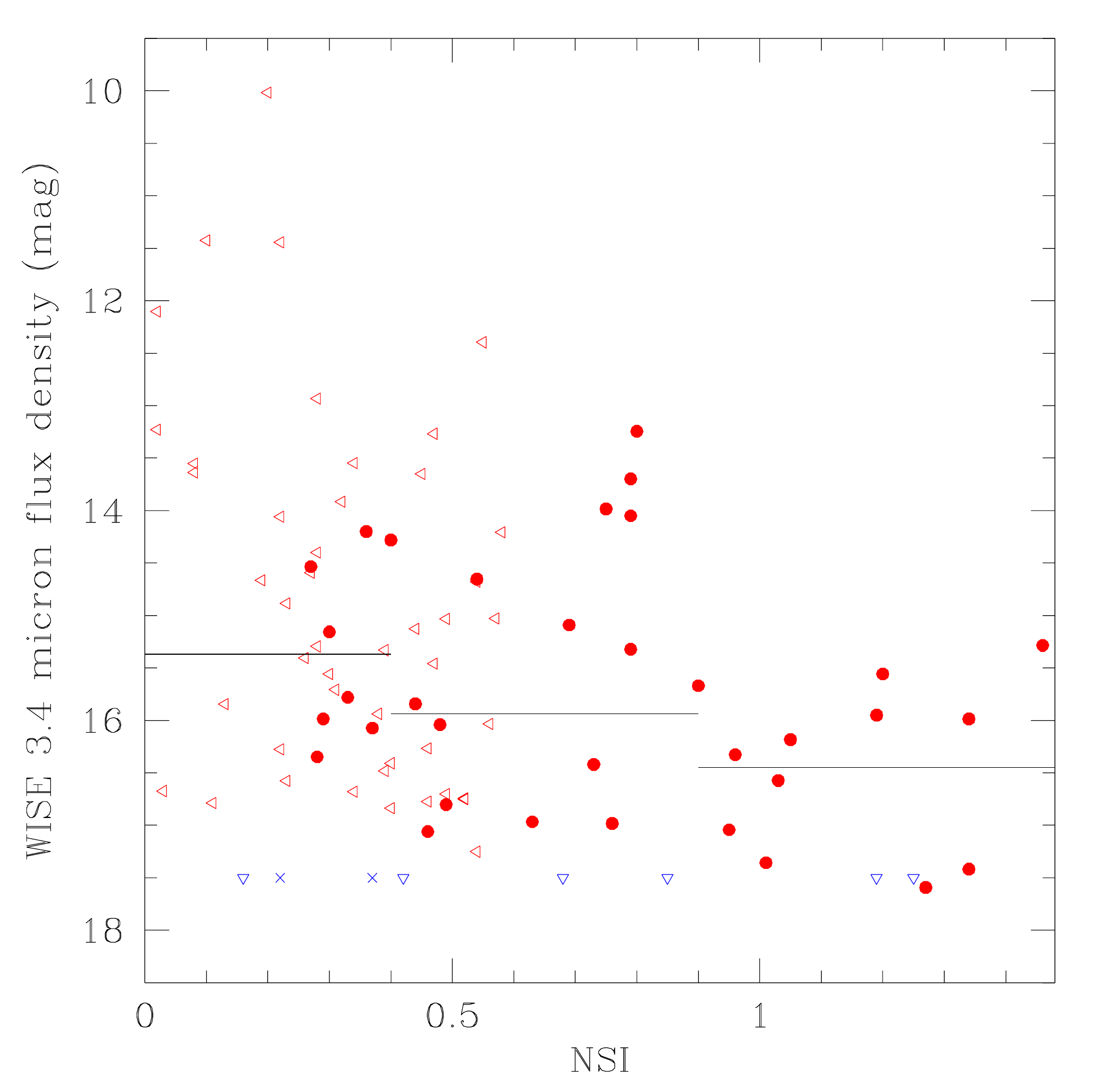}
  \caption{WISE W1 (3.4\,$\mu$m) magnitude plotted against normalised scintillation index (NSI) for all objects in the main sample (Table \ref{tab:data2}). Red points show radio sources matched with a WISE counterpart - filled red points show objects with an IPS detection, while open triangles represent IPS non-detections with an upper limit in NSI. 
Blue points show sources not matched with WISE, for which we assume W1 $<17.5$ mag. Blue triangles show objects with IPS detections, and blue crosses objects with an upper limit in both W1 and NSI. The three horizontal black lines show the median W1 magnitudes for the three scintillation classes discussed in \S2.3 of the text. }
    \label{fig:rajan_88c}
\end{figure}

Table \ref{tab:main_wise} compares the WISE properties of main-sample sources in the three scintillation classes defined by \cite{Chhetri2018}. As can also be seen from Figure \ref{fig:rajan_88c}, the compact sources with NSI $\geq$0.9 are typically significantly fainter in W1 than the extended sources with NSI $<$0.9. There are at least two plausible reasons for this - (i) the compact sources could be on average more distant than the extended sources, or (ii) the compact sub-sample could contain fewer QSOs (which are generally brighter in W1 than radio galaxies) than the extended sample. We consider this further in \S3.2, where we discuss the relationship between W1 and redshift for the radio galaxy and QSO populations. 

\begin{table}
 \setcounter{table}{5}
	\centering
	\caption{Overview of the WISE properties of sources in the main IPS-FIRST sample (Table \ref{tab:data2}). }
	\label{tab:main_wise}
\begin{tabular}{lcccccccll} 
\hline
IPS & n & Fraction w. & Median W1  \\
    &   & WISE ID     & (mag)   \\
\hline
All sources &  88 & 91\%    & 15.81 \\
            &     & (80/88) & $\pm0.21$   \\
& \\
Weak/no scint. & 40 & 93\% & 15.37  &  \\
(NSI $<$ 0.4) & & (37/40) & $\pm0.35$ \\
 & \\
Moderate scint. & 18 & 83\% & 15.94  & \\
(0.4$\leq$ NSI $<$ 0.9) & & (15/18)  & $\pm0.43$  & 
 & \\
 & \\
 Strong scint. & 14  & 86\% & 16.45 &  \\ 
(all w. NSI $\geq$ 0.9) &  & (12/14) &  $\pm0.27$ &   \\
\hline
	\end{tabular}
\end{table}

\subsubsection{The expanded compact source sample}

\begin{table}
	\centering
	\caption{Overview of the WISE properties of sources in the expanded compact sample (Table \ref{tab:data3}). }
	\label{tab:compact_wise}
\begin{tabular}{lcccccccll} 
\hline
IPS & n & Fraction w. & Median W1 &  \\
    &   & WISE ID     & (mag) &  \\
\hline
All sources & 65  & 91\% & 16.33 &  \\ 
(NSI $\geq$ 0.9) &  & (59/65) &  $\pm0.19$ &   \\
& \\
Peaked-spectrum & 27 & 89\% & 16.66 & \\
\cite{Callingham2017} &  & (24/27) & $\pm0.34$  \\
& \\
Compact steep spectrum & 33 & 91\% & 16.38  & \\
($\alpha<-0.5$) & & (30/33)  & $\pm0.25$  & \\
 & \\
 Flat-spectrum & 5 & 100\% & 15.86  &  \\
($\alpha>-0.5$) & & (5/5) & $\pm0.25$ \\
\hline
	\end{tabular}
\end{table}

Table \ref{tab:compact_wise} gives an overview of the WISE properties of the expanded compact sample from Table \ref{tab:data3} The overall ID rate for these sources (91\%) is similar to that for sources in the main IPS-FIRST sample (see Table \ref{tab:main_wise}). 

Within the expanded compact sample, we can identify three sub-classes based on the low-frequency radio spectrum and 162-1400\,MHz spectral index. 

\begin{itemize}
\item
{\bf Compact steep-spectrum} (CSS) sources with power-law radio spectra and 162-1400\,MHz spectral index $\alpha<-0.5$. 33 (51\%) of the strongly-scintillating sources fall into this class 
\item 
{\bf Peaked-spectrum sources: } 27 (42\%) of the strongly-scintillating (NSI$\geq0.9$) sources in the expanded compact sample are peaked-spectrum objects identified by \cite{Callingham2017}. As noted by \cite{Chhetri2018}, a surprisingly high fraction of all compact low-frequency sources have peaked radio spectra. 
\item
{\bf Compact flat-spectrum sources: } Only 7\% of the strongly-scintillating sources have `flat' radio spectra with 162-1400\,MHz spectral index $\alpha\geq-0.5$. These flat-spectrum objects are a minority population and (as can be seen from Table \ref{tab:compact_wise} they are significantly brighter in W1 than the other compact sources. Three of the five objects in this class are known to be QSOs. The other two have no optical spectrum currently available. 
\end{itemize}

Both the compact steep-spectrum and peaked-spectrum sub-samples contain many objects with faint W1 magnitudes. The median W1 magnitudes for both classes are similar - implying that we see a genuine distinction between a majority population of steep- and peaked-spectrum compact sources associated with (mainly) faint galaxies, and a smaller population of flat-spectrum sources associated mainly with radio-loud QSOs. 
Our current sample has only two strongly-scintillating sources with $\alpha<-1$ (i.e. `ultra-steep spectrum' sources as defined by \cite{Jarvis2001a}), but we note that only four of the seven compact steep-spectrum sources with $\alpha<-0.9$ (57\%) have a WISE ID, suggesting that many strongly-scintillating sources with very steep spectra may lie at high redshift. A larger sample is clearly needed to test this further. 
An investigation of the radio source counts for different sub-populations of strongly scintillating sources is presented in a companion paper by \cite{Chhetri2018b}; Paper III in this series. 

\subsubsection{The IPS-MRC-1\,Jy sample}
Table \ref{tab:main_mrc1jy} summarises the properties of the hosts of radio sources in the IPS-MRC-1Jy sample, 84\% of which are classified as galaxies and 16\% as radio-loud QSOs. 
It is notable that of the 14 QSOs in the sample, only one (MRC 0040-208 at z=0.655) is a strongly scintillating source with NSI $\geq0.9$. 

The MRC-1Jy sample can be reliably divided into sub-samples of radio galaxies and QSOs, so we can now address the issue raised in \S 3.1.2 of whether the median W1 magnitude of strongly-scintillating sources is fainter because the objects are more distant, or because the samples contain a different mix of (fainter) galaxies and (brighter) QSOs. 
Table \ref{tab:mrc_gal} lists the median W1 magnitude and 162-1400\,MHz spectral index for the radio galaxies in  the MRC sample, with QSOs excluded. 

The $\sim$1.5 mag difference in median W1 magnitude between the hosts of weakly-scintillating and strongly-scintillating sources remains when QSOs are excluded from the calculation. From this, we conclude that the difference is not due to a different mix of (brighter) QSOs and (fainter) galaxies as suggested in \S3.1.2. Instead, it must arise either because the strongly-scintillating sources are more distant or because their host galaxies are less luminous at mid-IR wavelengths. To resolve this, we now consider what is known about the redshift distribution of the IPS sources.  

\begin{table}
	\centering
	\caption{Overview of the WISE properties of sources in the IPS-MRC-1\,Jy sample (Table \ref{tab:data4}). }
	\label{tab:main_mrc1jy}
\begin{tabular}{lcccccccll} 
\hline
IPS & n & Fraction w. & \multicolumn{1}{c}{Median W1}  \\
    &   & WISE ID     & mag & \\
 \hline
All sources &  85 & 95\%    & 15.45 \\
            &     & (81/85) & $\pm0.25$   \\
& \\
Weak/no scint. & 53 & 96\% & 15.10  &  \\
(NSI $<$ 0.4) & & (51/53) & $\pm0.34$ \\
Moderate scint. & 18 & 94\% & 15.68  & \\
(0.4$\leq$ NSI $<$ 0.9) & & (17/18)  & $\pm0.44$  & \\
 Strong scint. & 8  & 88\% & 16.64 &  \\ 
(all w. NSI $\geq$ 0.9) &  & (7/8) &  $\pm0.81$ & \\
& \\
All QSOs & 14 & 100\% & 14.16 & \\ 
\citep{Kapahi1998b}  &  & (14/14) &  $\pm0.43$ & \\%
All galaxies & 71  & 88\% & 15.56 & \\ 
\citep{McCarthy1996}  &  & (67/71) &  $\pm0.28$ & \\ 
\hline
	\end{tabular}
\end{table}

\begin{table}
	\centering
	\caption{Median W1 magnitude and 162-1400\,MHz radio spectral index for galaxies in the IPS-MRC-1\,Jy sample (with QSOs excluded). }
	\label{tab:mrc_gal}
\begin{tabular}{lcccccccll} 
\hline
IPS & n &  \multicolumn{2}{c}{Median}  &  \\
    &   & W1 & $\alpha$ \\
    &   & mag & \\ 
 \hline
Weak/no scint. & 43 & 15.33  & -0.78 \\
(NSI $<$ 0.4)  &    & $\pm0.41$ & $\pm0.05$ \\
 & \\
Moderate scint. & 15 & 15.79 & -0.80 & \\
(0.4$\leq$ NSI $<$ 0.9) & & $\pm0.33$ & $\pm0.05$\\
 & \\
 Strong scint. & 7  & 16.92 & -0.78 \\ 
(all w. NSI $\geq$ 0.9) &  & $\pm0.53$ & $\pm0.13$  \\
\hline
	\end{tabular}
\end{table}

\subsection{Estimated redshift distribution for the compact low-frequency sources}
The K-$z$ relation between the near-infrared K-band magnitude and redshift of powerful radio galaxies \citep[e.g.][]{Willott2003} is commonly used to estimate the redshift of radio galaxies if a spectroscopic redshift measurement is not available. For the large-area \cite{Chhetri2018} IPS field, mid-infrared data from the WISE survey are significantly deeper than the available K-band data from the 2MASS survey, so we investigated the use of the WISE W1 magnitude as an alternative redshift indicator.  Like the 2.2$\mu$m K-band, the 3.4$\mu$m WISE W1 band mainly tracks the light of the stellar population in normal galaxies, so there is a reasonable justification for the use of the W1 magnitude as a photometric redshift estimator for radio galaxies in the same way that K-band \citep{Willott2003} and $r$-band \citep{Burgess2006b} have previously been used. 

The observed K--W1 colour of a galaxy will change significantly with redshift, due to differing k-corrections in the K and W1 bands \citep{Jarrett2017}, so making a direct conversion from the well-established K-$z$ relation to a new W1-$z$ relation is not straightforward. Instead, we chose to derive a simple empirical relation between W1 and redshift that we can use to estimate the likely redshift distribution of the strongly-scintillating radio sources in the \cite{Chhetri2018} IPS field. 

\subsubsection{The W1-z relation for radio galaxies}

\begin{figure}
	\includegraphics[width=\columnwidth]{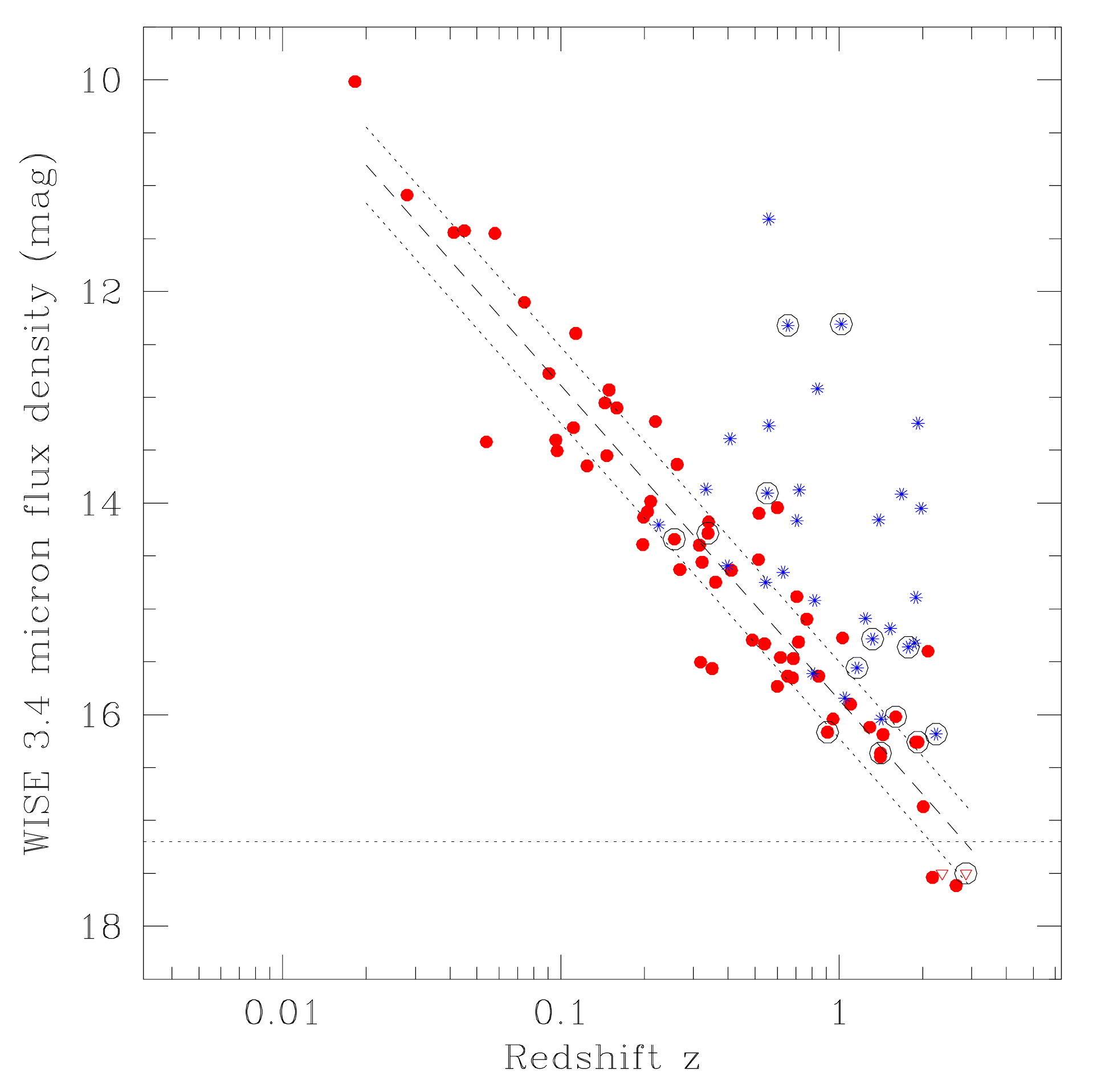}
  \caption{WISE W1 (3.4\,$\mu$m) magnitude plotted against redshift for objects in Tables \ref{tab:data2}, \ref{tab:data3} or \ref{tab:data4} that have a reliable spectroscopic redshift available in the literature. Red points show galaxies, and blue points QSOs. The dotted horizontal line at W1 = 17.2\,mag shows the completeness limit of the WISE catalogue, and the two red triangles below this line are upper limits in W1 for objects not detected in WISE. Open black circles show compact sources with strong IPS (NSI $\geq$ 0.9). The dashed line is a simple linear fit to the data points for galaxies, as discussed in the text. Offsets of $\pm0.36$\,mag from this line are also shown, and correspond to the median scatter of galaxies about the fitted line. }
    \label{fig:rajan_z1}
\end{figure}

Figure \ref{fig:rajan_z1} plots WISE W1 magnitude against spectroscopic redshift for objects in Tables \ref{tab:data2}, \ref{tab:data3} and \ref{tab:data4} that have a reliable redshift measurement in the literature. A limit of W1 $>17.5$ mag is assumed for the small number of sources with no catalogued W1 magnitude. 

Strongly-scintillating sources with known redshifts are marked by open circles in Figure \ref{fig:rajan_z1}. The red points for galaxy hosts of strongly-scintillating sources follow the same W1-$z$ relation as the host galaxies of other sources in the field. 
From this, we conclude that the fainter W1 magnitudes found for strongly-scintillating sources in \S3.1.4 (see Table \ref{tab:mrc_gal}) arise mainly because the strongly-scintillating sources are more distant, and not because their host galaxies are less luminous than those of other bright low-frequency radio sources. 

The dashed line in Figure \ref{fig:rajan_z1} shows the relation 
\begin{equation}
{\rm W1} = 15.860 + 2.976\,{\rm log(z)}, 
\end{equation}
which represents a simple linear fit to the galaxy points in Figure \ref{fig:rajan_z1}. The results shown here imply that powerful radio galaxies with W1 $>$ 16.8\,mag are likely to lie at redshift $z>2$. 
This appears reasonably consistent with our expectations from the \cite{Willott2003} K-$z$ relation:  
\begin{equation}
{\rm K} = 17.37 + 4.53\,{\rm log(z)} -0.31\,{\rm log(z)^2}, 
\end{equation}
given the strong redshift dependence of the observed (K-W1) colour due to band-shifting effects \citep[see Figure 8b of][]{Jarrett2017}.

\subsubsection{Redshift estimates for the strongly-scintillating IPS sources}
We now use the W1-$z$ relation from equation (1) to estimate the likely redshift distribution for the strongly-scintillating sources in the \cite{Chhetri2018} IPS field. 
We caution that that these redshift estimates are intended to be indicative only. In particular, the sources in the IPS sample are a mixture of galaxies and QSOs and so a W1-$z$  relation based on galaxy data only  provides a lower limit to the redshift of an individual radio source of unknown type. 

Table \ref{tab:ips_W1z} summarizes the results. The estimated median redshift for all sources in the main IPS-FIRST sample is $z\sim1.0$, roughly consistent with what is expected for sources in a survey of this depth at 162\,MHz \citep[e.g.][]{Condon1998}. 20\% of these sources have a host galaxy with W1 $>16.8$ mag, and are expected to be high-redshift radio galaxies (HzRGs) at redshift $z>2$. 
Sources that show weak or no scintillation have a significantly lower median redshift than the sample as a whole. 

\begin{table}
	\centering
	\caption{Estimated median redshifts for different source classes. The final column lists the fraction of sources with W1 $>16.8$\,mag, i.e. sources for which the estimated redshift is $z>2$.  }
	\label{tab:ips_W1z}
\begin{tabular}{lrrcrcccll} 
\hline
IPS & n &  \multicolumn{1}{c}{median} & \multicolumn{1}{c} {median} & W1 \\ 
    &   & W1 mag & $z\_{\rm est}$ &  $>16.8$ \\
 \hline
\multicolumn{3}{l}{\bf Main IPS-FIRST sample} \\
All sources & 88 & 15.81 &  0.96 &  20\% \\
            &    & $\pm0.21$ & $\pm0.18$ &  \\
Weak/no scint. & 40 & 15.37  & 0.68 & 8\%  \\
(NSI $<$ 0.4)  &    & $\pm0.35$ & $\pm0.20$  & \\
& \\
\multicolumn{3}{l}{\bf Expanded compact sample} \\
All with NSI$ \geq0.9$ & 65 & 16.33 & 1.44 & 37\%\\
                       &    & $\pm0.19$ & $\pm0.26$ & \\
Peaked-spectrum & 27 & 16.66 & 1.86 & 48\%  \\
                &    & $\pm0.34$ & $\pm0.53$ &  \\
Compact steep spectrum & 33 & 16.38 & 1.50 & 33\% \\ 
                       &    & $\pm0.25$ & $\pm0.33$ & \\
& \\
\multicolumn{3}{l}{\bf IPS-MRC-1Jy sample} \\
All galaxies & 71 & 15.56 & 0.79 & 17\% \\
             &    & $\pm0.28$ & $\pm0.18$ \\
Galaxies w. redshifts & 41 & 15.28 & 0.64 & 12\% \\
                      &    & $\pm0.39$ & $\pm0.21$  \\
Galaxies w. redshifts      &   \multicolumn{4}{l}{Median $z_{\rm spec}$ =  0.60} \\
 Galaxies w. redshifts      &  \multicolumn{4}{l}{Fraction with $z_{\rm spec}>2$ = 15\%} \\
\hline
\end{tabular}
\end{table}

The strongly-scintillating sources in the expanded compact sample are significantly more distant than the overall low-frequency source population, with an estimated median redshift of $z\sim1.5$. At least  one-third of all these sources have hosts that are fainter than 16.8\,mag in W1, and so are expected to lie at $z>2$. 

Finally, if we examine the subset of the IPS-MRC-1Jy sample for which reliable spectroscopic redshifts are available, we see a reassuring consistency between the W1-based estimates and spectroscopic measurements of both the median redshift and the fraction of sources likely to lie at $z>2$. The former is probably to be expected, since MRC spectroscopic redshifts were used (along with other data) to derive the W1-$z$ relation presented in \S3.2.1. However, the agreement between the estimated and actual fraction of MRC-1Jy galaxies at redshift $z>2$, both of which represent the tail of a distribution, is encouraging and suggests that IPS might potentially be used as a new tool to identify high-redshift radio galaxies (HzRGs) in the distant Universe. We explore this idea further in the next section. 

\section{IPS as a tool for selecting high-redshift radio galaxies? }

\subsection{An IPS-based selection method for HzRGs}
Current methods for identifying high-redshift radio galaxies (HzRGs) at $z>2$ generally involve the selection of sources with ultra-steep (USS) radio spectra (e.g. spectral index $\alpha<-1.3$; \cite{Miley2008}). Secondary criteria related to the angular size of the radio source and faintness of the host galaxy are also applied. Studies based on this technique \citep{vanBreugel1999, Jarvis2001b, DeBreuck2002, Bryant2009, Saxena2018} have been effective in selecting distant objects, but do not necessarily give a complete and unbiased picture of the high-redshift radio-source population. For example, most of the distant objects selected through the USS technique have radio emission extending over tens of kiloparsecs \citep{Miley2008}, suggesting that this technique may bias against the selection of young, compact radio galaxies. 

An IPS-based technique for selecting HzRG candidates could provide a useful alternative to traditional USS methods.
In particular: 
\begin{itemize}
\item 
Selecting strongly-scintillating sources dominated by a compact (sub-arcsec) radio component simplifies the process of identifying the host galaxy. For example, \cite{McCarthy1996} note that identification of the hosts of compact (angular size $\leq5$\,arcsec) and unresolved radio sources in their MRC-1Jy sample was straightforward, whereas more extended sources sometimes had two or more plausible identifications for which follow-up spectra were needed.  
\item
The IPS technique typically selects sources with angular size $\leq0.5$\,arcsec, corresponding to linear scales smaller than 3--5\,kpc at redshift $z>2$. In other words, this technique should select objects with linear size similar to classical GPS and CSS radio galaxies \citep{ODea1998}. While this is only a transient stage in the evolution of an individual radio galaxy, it is a stage that all powerful radio galaxies are expected to pass through -- allowing us to connect objects in the same evolutionary phase at different redshifts.  Since the USS technique often picks up more extended lobe-dominated radio galaxies, the USS and IPS techniques have the potential to be complementary in building up complete samples of high-redshift radio sources. 
\item
The median estimated redshift of $z\sim1.5$ for the compact sources in Table \ref{tab:ips_W1z} is not far below the median value of $z\sim1.9$ for the \cite{Jarvis2001a} 6C* sample, which used USS spectral-index selection (with $\alpha<-0.98$ at 151-4850\,MHz). The IPS technique may therefore be able to reach a comparable redshift depth to previous USS searches, but without the need to impose any spectral-index selection.  
\end{itemize}

An IPS-based technique for identifying HzRG candidates using data from a wide-field IPS survey with the MWA would be a fairly straightforward process: 
\begin{enumerate}
\item 
Select strongly-scintillating radio sources using an appropriate cutoff in the normalised scintillation index. 
\item 
Cross-match the NVSS or SUMSS radio positions of these sources with the allWISE mid-IR survey catalogue, and select objects fainter than some appropriate limit in W1 magnitude. 
\end{enumerate}

In the next section, we use the IPS-MRC-1Jy sample to make some first tests of the effectiveness and completeness of this IPS-based HzRG search technique. 

\subsection{Tests on the MRC-1Jy sample}
For a HzRG search technique to be effective, it should ideally perform well against two measures:
\begin{enumerate}
\item 
{\bf Efficiency: } What fraction of the selected candidates 
are in the desired redshift range (e.g. what fraction are at $z>2$)? 
\item
{\bf Completeness: } What fraction of all sources in the target redshift range are selected by the technique?
\end{enumerate}

The IPS-MRC-1Jy sample contains six $z>2$ radio AGN that were selected from a flux-limited survey without any spectral-index preselection, so we can use this sample to test the efficiency and completeness of the proposed IPS technique. Since this $z>2$ sample is very small, the results are only indicative at this stage. With IPS observations of the full MRC-1Jy survey area, more detailed tests would be possible. 

As a starting point, we assume the two selection criteria to be NSI $>0.8$, and W1 $>17.2$ mag. We chose this slightly lower cutoff in NSI partly based on the distribution of points in Figure \ref{fig:rajan_88c}, but also to avoid reducing the size of our small test sample even further. 
The results were as follows: 
\begin{enumerate}
\item 
{IPS Efficiency:} Of the 55 sources in Table \ref{tab:data4} with reliable redshift data, only two (MRC 0030-219 and MRC 0052-241) satisfy both selection criteria. These two objects both lie at $z>2$ (z = 2.168 and 2.86 respectively). Thus the IPS technique appears to be very efficient based on this small sample.  
\item 
{IPS Completeness: }
Of the six $z>2$ radio galaxies in the IPS-MRC-1Jy sample, two were identified by the IPS-based selection criteria, giving a completeness of $\sim$30\%. 
\end{enumerate}

We can make similar tests for the USS-based selection techniques used by \cite{Jarvis2001b} and \cite{Saxena2018}.  
\begin{enumerate}
\item 
{USS Efficiency: } 
Three of the 55 sources with redshifts in Table \ref{tab:data4} (MRC 0015-229, MRC 0052-241 and MRC 0140-257) satisfied the \cite{Jarvis2001b} selection criterion of $\alpha<-0.98$.
All three objects lie at $z>2$ (z = 2.01, 2.86 and 2.64 respectively). Thus the \cite{Jarvis2001b} USS technique also appears to be very efficient.  No source in Table \ref{tab:data4} has a steep enough radio spectrum to satisfy the \cite{Saxena2018} cutoff of $\alpha<-1.3$. 
\item
{USS Completeness:} 
Of the six $z>2$ radio galaxies in the IPS-MRC-1Jy sample, three satisfied the \cite{Jarvis2001b} selection criterion of $\alpha<-0.98$ giving a completeness of $\sim$50\%. None of the six $z>2$ sources satisfied the \cite{Saxena2018} selection criteria (which were designed for use with much deeper radio surveys), so the completeness for this more stringent USS cutoff of $\alpha<-1.3$ is $<17$\%. 
\end{enumerate}

Our conclusion from these tests (on an admittedly small sample) is that the IPS selection technique appears competitive with existing USS techniques in terms of efficiency and completeness. The IPS technique also has the advantage of being able to select high-redshift radio AGN candidates without any need for spectral-index preselection. 
Extending the work presented here to a larger IPS sample with additional redshift data would allow the performance of this technique to be tested in more detail. 

\section{Conclusions and future work}
In this paper, we have identified the hosts of a representative sample of radio sources for which Interplanetary Scintillation (IPS) measurements have been made by \cite{Chhetri2018}. We have also shown that WISE mid-IR magnitudes can provide a useful redshift estimator for these powerful radio galaxies, using only a simple linear relation between W1 and log($z$) (see Figure \ref{fig:rajan_z1}). 

The hosts of sub-arcsecond compact sources identified through their strong scintillation in MWA IPS observations are typically distant radio galaxies (median redshift $z\sim1.5$), 
and we estimate that over 30\% of them lie at redshift $z>2$. 
More follow-up optical/IR data, and a larger-area IPS survey are needed to take this work further, but the results to date look very encouraging. 

In the near future, the upgraded capabilities of the MWA should allow wide-field IPS studies to probe significantly deeper in flux density - reaching a typical detection limit of 0.4\,Jy at 162\,MHz, 
and detecting sources as weak as 0.1 Jy under ideal conditions (\cite{Morgan2018a}; Paper V in this series).  This should allow us to probe to even higher redshift, and potentially detect powerful compact radio galaxies and radio-loud QSOs out to redshift 5-6. 

\section*{Acknowledgements}

This scientific work makes use of the Murchison Radio-astronomy Observatory, operated by CSIRO. We acknowledge the Wajarri Yamatji people as the traditional owners of the Observatory site. Support for the operation of the MWA is provided by the Australian Government (NCRIS), under a contract to Curtin University administered by Astronomy Australia Limited. We acknowledge the Pawsey Supercomputing Centre which is supported by the Western Australian and Australian Governments.

Parts of this research were conducted by the Australian Research Council Centre of Excellence for All-sky Astrophysics (CAASTRO), through project number CE110001020. 

This publication makes use of data products from the Wide-field Infrared Survey Explorer, which is a joint project of the University of California, Los Angeles, and the Jet Propulsion Laboratory/California Institute of Technology, funded by the National Aeronautics and Space Administration.

Funding for the Sloan Digital Sky Survey IV has been provided by the Alfred P. Sloan Foundation, the U.S. Department of Energy Office of Science, and the Participating Institutions. SDSS-IV acknowledges
support and resources from the Center for High-Performance Computing at the University of Utah. The SDSS web site is www.sdss.org.

SDSS-IV is managed by the Astrophysical Research Consortium for the 
Participating Institutions of the SDSS Collaboration including the 
Brazilian Participation Group, the Carnegie Institution for Science, 
Carnegie Mellon University, the Chilean Participation Group, the French Participation Group, Harvard-Smithsonian Center for Astrophysics, 
Instituto de Astrof\'isica de Canarias, The Johns Hopkins University, 
Kavli Institute for the Physics and Mathematics of the Universe (IPMU) / 
University of Tokyo, Lawrence Berkeley National Laboratory, 
Leibniz Institut f\"ur Astrophysik Potsdam (AIP),  
Max-Planck-Institut f\"ur Astronomie (MPIA Heidelberg), 
Max-Planck-Institut f\"ur Astrophysik (MPA Garching), 
Max-Planck-Institut f\"ur Extraterrestrische Physik (MPE), 
National Astronomical Observatories of China, New Mexico State University, 
New York University, University of Notre Dame, 
Observat\'ario Nacional / MCTI, The Ohio State University, 
Pennsylvania State University, Shanghai Astronomical Observatory, 
United Kingdom Participation Group,
Universidad Nacional Aut\'onoma de M\'exico, University of Arizona, 
University of Colorado Boulder, University of Oxford, University of Portsmouth, 
University of Utah, University of Virginia, University of Washington, University of Wisconsin, 
Vanderbilt University, and Yale University.




\bibliographystyle{mnras}
\bibliography{references-IPS-new} 


\begin{landscape}
 \begin{table}
  \setcounter{table}{2}
  \caption{The main IPS-FIRST sample of high S/N sources from the Chhetri et al.\ (2018a) IPS sample that lie within the FIRST overlap area. }
  \label{tab:data2}
  \begin{tabular}{llrrrrclllllllll}
    \hline
    GLEAM name & Alt name & S$_{162}$ & $\pm$ & Norm & NSI & FIRST & Radio & \multicolumn{2}{c}{Radio position (J2000)} & WISE & W1 & $\pm$ & z  & z$\_$ref &Notes \\
     & & Jy & Jy & SNR & & comp & class & RA & Dec & match & mag  \\
 (1) & (2) & (3) & (4) & (5) & (6) & (7) & (8) & (9) & (10) & (11) & (12) & (13) & (14) & (15) & (16) \\
\hline
\multicolumn{10}{l}{{\bf (a) All high S/N sources in the FIRST survey area} (NormSNR $\geq12.5$) } \\
J000057-105435 & PKS 2358-111  & 4.775 & 0.023 & 23.1 & 0.63 & 1 & Single & 00 00 57.662 & -10 54 32.16 &  J000057.57-105432.1  & 16.966 & 0.134 & .. & .. & \\ 
J000829-055839 & 3C 003   & 7.643 & 0.018 & 27.6 & 0.79 & 1 & Single  & 00 08 29.354 & -05 58 45.91 &  J000829.30-055845.2  & 13.698 & 0.026 & .. & .. &  \\
J001356-091952 & PKS 0011-096   & 4.131 & 0.020 & 25.9 & $<0.28$ & 4 &  Double & 00 13 56.86 & -09 19 52.59 &  J001357.24-091949.5   & 15.295 & 0.041 & 0.4878 & SDSS & $\bigstar$ A \\ 
J001454-095911 & PKS 0012-102  & 2.136 & 0.020 & 16.0 & $<0.31$ & 2 & Double & 00 14 54.91  & -09 59 12.8  &  J001454.97-095912.2  & 15.708 & 0.048 & .. & .. & \\
J001821-111139 & MRC 0015-114  & 1.932 & 0.020 & 16.2 & 0.95 & 1 & Single & 00 18 22.052 & -11 11 38.87 &  J001821.92-111139.1  & 17.042 & 0.128 & .. & .. & \\
 & \\
J001859-102248 & PKS 0016-10   & 7.648 & 0.020 & 57.8 & 0.44 & 2 & Double & 00 18 59.48  & -10 22 49.2 &  J001859.52-102245.6   & 15.843 & 0.053 & .. &  .. & \\
J001931-043543 & 3C 007  & 4.523 & 0.018 & 16.4 & 1.03 & 1 & Single & 00 19 31.513 & -04 35 47.07 &  J001931.48-043548.1  & 16.573 & 0.087 & .. & ..  & \\
J002050-085731 & PKS 0018-09   & 6.327 & 0.019 & 42.3 & 0.29 & 2 & Double & 00 20 50.33  & -08 57 34.9 &  J002050.13-085733.5 & 15.986 & 0.062 & .. & .. & \\
J002208-104133 & PKS 0019-109  & 1.774 & 0.019 & 14.3 & 1.34 & 1 & Single & 00 22 09.010 & -10 41 33.29 &  J002209.00-104132.1  & 17.419 & 0.175 & .. & .. & P \\
J002223-070230 & PKS 0019-073    & 3.821 & 0.016 & 23.1 & 1.25 & 1 & Single & 00 22 23.050 & -07 02 35.23 & .. & .. & ..  & ..  & .. & $\bigstar$ P \\
& \\
J002319-074445 & PKS 0020-08   & 5.470 & 0.017 & 32.9 & 0.27 & 1 & Single (ext) & 00 23 19.031 & -07 44 49.69 &  J002319.02-074449.3  & 14.534 & 0.031 & 0.514 & SDSS & A \\ 
J002533-064253 & MRC 0023-069  & 2.226 & 0.015 & 15.1 & $<0.03$ & 3 & Triple & 00 25 33.751  & -06 43 02.19 &  J002533.77-064301.0  & 16.673 & 0.095 & .. & .. & \\
J002631-084234 & PKS 0024-089   & 2.253 & 0.019 & 16.4 & $<0.22$ & 2 & Double & 00 26 31.60  & -08 42 37.0 &  J002631.46-084235.5  & 14.058 & 0.029 & .. & .. & A \\ 
J002922-111151 & PKS 0026-114    & 3.822 & 0.018 & 38.8 & 0.73 & 1 & Single & 00 29 22.615 & -11 11 51.28 &  J002922.68-111150.9  & 16.420 & 0.075 & .. & .. & \\
J003102-091157 & PKS 0028-094    & 1.963 & 0.020 & 15.7 & $<0.40$ & 2 & Double & 00 31 03.00  & -09 12 04.2 &  J003102.89-091202.4  & 16.408 & 0.075 & .. & .. & \\
& \\
J003201-085133 & PKS 0029-091 & 2.125 & 0.020 & 16.6 & 1.27 & 1 & Single & 00 32 00.977 & -08 51 37.48 &  J003200.97-085139.1  & 17.593 & 0.212 & .. & .. & P \\
J003217-034903 & PKS 0029-040  & 3.779 & 0.017 & 15.2 & $<0.34$ & 2 & Double & 00 32 17.32  & -03 49 08.4 &  J003217.22-034909.1  & 13.548 & 0.030 & .. & .. & \\
J003251-101801 & MRC 0030-105  & 1.659 & 0.019 & 14.5 & $<0.37$ & 2 & Double & 00 32 51.91  & -10 18 05.0 & .. & .. &  & .. & .. & $\bigstar$ \\
J003354-073019 & PKS 0031-07 & 6.134 & 0.017 & 47.8 & 0.33 & 2 & Double & 00 33 54.66  & -07 30 28.6 &  J003354.74-073029.7  & 15.781 & 0.051 & .. & .. & \\
J003502-083437 & MRC 0032-088  & 2.360 & 0.019 & 21.2 & 0.36 & 3 & Triple & 00 35 02.72 & -08 34 37.38 &  J003503.45-083440.1   & 14.200 & 0.029 & .. & .. & \\
 & \\
J003616-111650 & PKS 0033-115 & 1.837 & 0.019 & 21.0 & 0.30 & 4 & Double & 00 36 16.81 & -11 16 50.58 &  J003615.94-111657.8   & 15.157 & 0.041 & .. & .. & \\
J003704-010910 & 3C 015 & 19.710 & 0.024 & 35.8 & $<0.02$ & 3 & Triple & 00 37 03.959 & -01 09 04.45 &  J003704.10-010908.2   & 12.103 & 0.024 & 0.0737 & SDSS & A \\
J003820-020737 & 3C 017 & 28.389 & 0.026 & 60.7 & $<0.02$ & 3 & Triple & 00 38 20.30 & -02 07 37.5 &  J003820.53-020740.5   & 13.228 & 0.024 & 0.2204 & SDSS & A \\ 
J003822-111244 & PMN J0038-1112 & 1.347 & 0.018 & 15.4 & $<0.39$ & 1 & Single (ext) & 00 38 22.781 & -11 12 48.27 &  J003822.80-111250.2  & 15.331 & 0.041 & .. & .. &  \\
J003924-080223 & MRC 0036-083  & 1.520 & 0.016 & 13.4 & $<0.47$ & 1 & Single  (ext) & 00 39 25.019 & -08 02 25.59 &  J003924.98-080227.0  & 15.459 & 0.043 & 0.6165 & SDSS &  \\
& \\
J003938-074610 & MRC 0037-080 & 2.493 & 0.017 & 23.1 & $<0.39$ & 1 & Single  (ext) & 00 39 38.245 & -07 46 10.74 &  J003938.31-074610.9  & 16.480 & 0.080 & ..  & .. & \\
J004049-094832 & MRC 0038-100 & 2.173 & 0.022 & 19.6 & 0.42 & 2 & Double & 00 40 48.56  & -09 48 30.9 & .. & .. & .. & .. & .. & $\bigstar$ \\
J004125-014314 & PKS 0038-019  & 7.811 & 0.021 & 17.5 & $<0.32$ & 2 & Double & 00 41 26.02  & -01 43 16.0 &  J004126.00-014315.7  & 13.915 & 0.027 & 1.679 & Du89 & A \\ 
J004232-081543 &  MRC 0040-085 & 2.296 & 0.018 & 21.3 & $<0.34$ & 2 & Double & 00 42 32.26  & -08 15 48.1 &  J004232.23-081548.8  & 16.680 & 0.090 & .. & .. & \\
J004246-061325 & PKS 0040-06 & 2.237 & 0.016 & 15.6 & $<0.45$ & 2 & Double & 00 42 46.84 & -06 13 53.0 & J004246.84-061353.0 & 13.649 & 0.025 & 0.1243 & SDSS & $\bigstar$ \\
 & \\
\hline
\end{tabular}
\end{table}
\end{landscape}

\newpage 

\begin {landscape}
\begin{table}
  \begin{tabular}{llrrrrclllllllll}
    \hline
    GLEAM name & Alt name & S$_{162}$ & $\pm$ & Norm & NSI & FIRST & Radio & \multicolumn{2}{c}{Radio position (J2000)} & WISE & W1 & $\pm$ & z  & z$\_$ref &Notes \\
     & & Jy & Jy & SNR & & comp & class & RA & Dec & match & mag  \\
 (1) & (2) & (3) & (4) & (5) & (6) & (7) & (8) & (9) & (10) & (11) & (12) & (13) & (14) & (15) & (16) \\
\hline
J004409-050659 & PKS 0041-053 & 2.652 & 0.016 & 17.2 & $<0.52$ & 1 & Single (ext) & 00 44 09.735 & -05 07 03.02 &  J004409.82-050702.5  & 16.744 & 0.104 & .. & .. &  \\
J004644-052157 & PKS 0044-05 & 2.933 & 0.016 & 21.2 & 0.79 & 1 & Single & 00 46 44.519 & -05 22 00.29 &  J004644.54-052200.0  & 15.323 & 0.041 & 1.869 & Wi76 &   \\
J004707-070451 & PKS 0044-073 & 2.435 & 0.016 & 19.4 & $<0.28$ & 4 & Double & 00 47 07.86 & -07 04 51.4 &  J004708.26-070445.4  & 14.400 & 0.029 & .. & .. & \\
J004820-092212 & MRC 0045-096   & 2.064 & 0.019 & 24.4 & 0.37 & 2 & Double & 00 48 20.65   & -09 22 14.8 &  J004820.59-092215.4  & 16.074 & 0.063 & .. & .. & \\
J004858-062832 & PKS 0046-06 & 3.059 & 0.015 & 25.8 & 0.40 & 2 & Double & 00 48 58.40  & -06 28 31.0 &  J004858.02-062831.0 & 14.282 & 0.028 & ..  & .. & \\
& \\
J004900-054358 & MRC 0046-060 & 1.519 & 0.015 & 14.3 & $<0.52$ & 1 & Single (ext) & 00 49 00.852 & -05 44 00.89 &  J004900.83-054400.5  & 16.752 & 0.097 & .. & .. & \\
J004945-024257 & PKS 0047-02 & 7.862 & 0.02 & 28.7 & $<0.11$ & 3 & Triple  & 00 49 45.824  & -02 42 59.99  &  J004945.33-024258.8   & 16.788 & 0.107 & .. & .. & \\
J004954-100613 & PKS 0047-10 & 2.257 & 0.018 & 27.3 & 1.05 & 1 & Single & 00 49 54.107 & -10 06 14.95 &  J004954.03-100614.5  & 16.183 & 0.061 & 2.236 & SDSS & A \\
J005039-102734 & MRC 0048-107 & 2.138 & 0.018 & 26.8 & 0.90 & 1 & Single & 00 50 39.004 & -10 27 35.77 & .. & 15.67 & 0.05 & .. & .. & $\bigstar$ \\
J005049-032219 & MRC 0048-036 & 3.230 & 0.020 & 17.0 & $<0.54$ & 2 & Double & 00 50 49.87  & -03 22 21.5 &  J005049.96-032222.2  & 17.251 & 0.163 & .. & .. & \\
& \\
J005108-065001 & PKS 0048-071  & 1.773 & 0.016 & 16.9 & 0.79 & 1 & Single & 00 51 08.208 & -06 50 02.25 &  J005108.20-065002.1  & 14.050 & 0.028 & 1.975 & Wr83 & AF  \\
J005234-093819 & MRC 0050-099 & 1.472 & 0.018 & 17.2 & 1.19 & 1 & Single & 00 52 34.551 & -09 38 18.78 & .. & .. & .. & ..  & .. & \\
J005354-092824 & MRC 0051-097  & 1.634 & 0.019 & 18.9 & $<0.08$ & 4 & Double & 00 53 54.14 & -09 28 24.1 &  J005354.20-092822.4  & 13.635 & 0.026 & 0.262 & SDSS &  \\
J005408-033354 & 3C 026 & 11.017 & 0.021 & 55.7 & 0.75 & 1 & Single  & 00 54 08.433 & -03 33 55.31 &  J005408.43-033355.2  & 13.984 & 0.028 & 0.2104 & Sc65 & A \\
J005447-091847 & PKS 0052-096   & 1.890 & 0.019 & 22.7 & $<0.26$ & 1 & Single (ext) & 00 54 47.525 & -09 18 48.04 &  J005447.42-091845.0  & 15.405 & 0.042 & .. & .. &  \\
 & \\
J005447-095257 & PMN J0054-0952  & 1.432 & 0.019 & 17.3 & $<0.22$ & 2 & Double & 00 54 47.81  & -09 52 54.6 &  J005447.87-095254.6  & 16.276 & 0.068 & .. & .. &   \\
J005536-095216 & MRC 0053-101  & 2.140 & 0.020 & 25.3 & $<0.22$ & 2 & Double & 00 55 36.69 & -09 52 13.9 & .. & .. & .. & .. & .. & $\bigstar$ \\
J005551-082956 & MRC 0053-087  & 1.534 & 0.017 & 16.4 & $<0.23$ & 2 & Double & 00 55 51.66  & -08 29 51.6 &  J005551.49-082953.8 & 14.885 & 0.033 & 0.7049 & SDSS &   \\
J005734-012329 & 3C 029 & 16.168 & 0.03 & 41.5 & $<0.10$ & 4 & Wide double & 00 57 34.92 & -01 23 27.5 & J005734.90-012327.5 & 11.425 & 0.022 & 0.0450 & We99 & $\bigstar$ A   \\
J005805-053952 & PKS 0055-059   & 1.994 & 0.016 & 14.8 & 0.69 & 1 & Single  & 00 58 05.070 & -05 39 52.45 &  J005805.08-053952.3  & 15.091 & 0.036 & 1.2456 & Ti11 & AF   \\
 &  \\
J005903-075959 & MRC 0056-082  & 1.355 & 0.016 & 16.9 & $<0.46$ & 2 & Double & 00 59 03.09  & -07 59 59.2 &  J005903.11-080001.0  & 16.267 & 0.068 & .. & .. &  \\
J005928-101733 & PMN J0059-1017  & 1.219 & 0.018 & 16.8 & 0.48 & 2 & Double & 00 59 29.10  & -10 17 34.5 &  J005929.11-101734.8  & 16.04 & 0.056 & 1.419 & SDSS &  \\
J010042-050237 & MRC 0058-053  & 1.91 & 0.016 & 13.2 & 1.34 & 1 & Single  & 01 00 42.675 & -05 02 36.84 &  J010042.64-050236.5  & 15.987 & 0.052 & .. & .. & P   \\
J010159-105555 & TN J0102-1055 & 1.774 & 0.019 & 23.3 & 0.85 & 1 & Single & 01 02 00.001 & -10 55 56.36 & .. & .. & .. & .. & .. & $\bigstar$  \\
J010509-065250 & PKS 0102-07  & 3.217 & 0.016 & 33.3 & $<0.23$ & 2 & Wide double & 01 05 09.44  & -06 52 56.8 &  J010509.55-065257.1  & 16.577 & 0.072 & .. & .. & \\
 &  \\
J010834-103346 & PKS 0106-108 & 4.38 & 0.019 & 56.1 & $<0.13$ & 2 & Wide double & 01 08 34.62  & -10 33 45.1  &  J010834.61-103345.7  & 15.844 & 0.048 & .. & .. &  \\
J010930-085303 & PKS 0106-091  & 2.048 & 0.018 & 26.7 & 0.46 & 2 & Double & 01 09 30.32  & -08 53 07.3 & J010930.32-085304.9  & 17.060 & 0.114 & .. & .. &   \\
J011212-045504 &  MRC 0109-051  & 2.531 & 0.016 & 16.4 & $<0.57$ & 3 & Triple & 01 12 13.495  & -04 55 10.76 &  J011213.45-045509.7  & 15.029 & 0.035 & .. & .. &  \\
J011242-080404 & MRC 0110-083  & 1.987 & 0.017 & 23.0 & $<0.19$ & 2 & Double & 01 12 42.50  & -08 04 02.3 &  J011242.72-080402.3 & 14.664 & 0.03 & .. & .. &  \\
J011312-101419 & PKS 0110-105  & 1.565 & 0.019 & 20.2 & 1.46 & 1 & Single & 01 13 12.101 & -10 14 21.57 &  J011312.09-101421.2  & 15.285 & 0.037 & 1.3210 & SDSS & A   \\
 & \\
J011406-100800 & PMN J0114-1007 & 1.135 & 0.020 & 15.4 & 0.49 & 1 & Single (ext) & 01 14 06.559 & -10 08 01.00 &  J011406.33-100802.6 & 16.802 & 0.089 & .. & .. & $\bigstar$ \\
J011547-085756 & PMN J0115-0857 & 1.076 & 0.019 & 13.3 & $<0.46$ & 3 & Triple & 01 15 47.404  & -08 57 56.46 &  J011547.46-085756.8  & 16.772 & 0.098 & .. & .. & A   \\
J011645-041849 & MRC 0114-045 & 2.007 & 0.017 & 12.6 & $<0.27$ & 1 & Single (ext) & 01 16 45.265 & -04 18 54.84 &  J011645.21-041855.8  & 14.593 & 0.032 & 0.3965 & SDSS & $\bigstar$  \\
J011751-100557 & PMN J0117-1006  & 1.125 & 0.020 & 13.2 & $<0.40$ & 2 & Double & 01 17 51.40  & -10 05 53.0 &  J011751.35-100555.6   & 16.836 & 0.105 & .. & .. &   \\
J011815-012037 & PKS 0115-01  & 5.357 & 0.020 & 16.4 & 1.20 & 1 & Single & 01 18 15.424 & -01 20 30.12 &  J011815.39-012030.2  & 15.557 & 0.047 & 1.162 & Ho03 & AP \\
 &  \\
\hline
\end{tabular}
\end{table}
\end{landscape}

\newpage 

\begin {landscape}
\begin{table}
  \begin{tabular}{llrrrrclllllllll}
    \hline
    GLEAM name & Alt name & S$_{162}$ & $\pm$ & Norm & NSI & FIRST & Radio & \multicolumn{2}{c}{Radio position (J2000)} & WISE & W1 & $\pm$ & z  & z$\_$ref & Notes \\
     & & Jy & Jy & SNR & & comp & class & RA & Dec & match & mag  \\
 (1) & (2) & (3) & (4) & (5) & (6) & (7) & (8) & (9) & (10) & (11) & (12) & (13) & (14) & (15) & (16) \\
\hline
J011931-102506 & TXS 0117-106 & 1.194 & 0.020 & 15.9 & $<0.49$ & 1 & Single & 01 19 31.320 & -10 25 08.72 &  J011931.25-102508.7  & 16.701 & 0.087 & .. & .. &  \\
J012227-042123 & PKS 0119-04  & 4.770 & 0.018 & 29.6 & 0.80 & 1 & Single & 01 22 27.921 & -04 21 27.19 & J012227.88-042127.1  & 13.245 & 0.025 & 1.925 & Os94 & $\bigstar$ AP   \\
J012231-061953 & MRC 0120-065  & 1.573 & 0.017 & 14.1 & $<0.49$ & 4 & Triple & 01 22 32.00 & -06 19 57.7 & J012232.66-061958.5  & 15.034 & 0.034 & .. & .. & $\bigstar$  \\
J012434-040101 & PKS 0122-042  & 3.128 & 0.018 & 17.9 & $<0.47$ & 3 & Triple  & 01 24 34.794  & -04 01 02.28 &  J012434.81-040102.7  & 13.267 & 0.024 & 0.5600 & SDSS & A   \\
J012603-012356 & NGC 547 & 6.089 & 0.029 & 16.4 & $<0.20$ & 2 & Complex & 01 26 00.32 & -01 20 42.6 & J012600.61-012042.4 & 10.016 & 0.022 & 0.0182 & RC3 & $\bigstar$ \\
 &  \\
J013029-083321 & PKS 0127-08  & 1.786 & 0.02 & 15.8 & $<0.44$ & 2 & Double & 01 30 29.87 & -08 33 19.9 &  J013029.94-083319.7  & 15.128 & 0.037 & .. & .. &  \\
J013136-070354 & PKS 0129-073  & 8.868 & 0.021 & 62.6 & 0.16 & 2 & Double & 01 31 36.45  & -07 03 57.3 & .. & .. & .. & .. & .. & $\bigstar$ \\
J013157-081949 & PKS 0129-085 & 2.183 & 0.02 & 17.8 & $<0.28$ & 3 & Triple & 01 31 57.572 & -08 19 58.96 &  J013157.78-081954.9  & 12.931 & 0.023 & 0.1491 & 6dFGS &  \\
J013212-065232 & PKS 0129-07 & 5.309 & 0.02 & 36.7 & 1.19 & 1 & Single & 01 32 12.181 & -06 52 35.92 &  J013212.24-065236.0  & 15.951 & 0.051 & .. & .. & AP   \\
J013325-102616 & PKS 0130-106  & 1.168 & 0.02 & 12.9 & $<0.55$ & 2 & Double & 01 33 25.92  & -10 26 17.3 &  J013325.87-102618.6  & 12.396 & 0.024 & 0.1132 & SDSS &  \\
  & \\
J013524-051630 & PKS 0132-055  & 2.048 & 0.017 & 14.0 & $<0.58$ & 7 & Wide triple & 01 35 23.741  & -05 16 29.81 &  J013523.71-051629.8  & 14.208 & 0.027 & 0.225 & SDSS &   \\
J013635-080604 & MRC  0134-083  & 2.644 & 0.021 & 18.4 & $<0.08$ & 2 & Double & 01 36 35.65  & -08 06 04.2 &  J013635.64-080608.6   & 13.551 & 0.053 & 0.1464 & 6dFGS &  \\
J013715-091141 & MRC 0134-094  & 2.618 & 0.021 & 25.2 & $<0.22$ & 4 & Complex & 01 37 15.81 & -09 11 41.8 &  J013715.38-091151.2   & 11.443 & 0.022 & 0.0413 & We99 &  \\
J013716-093105 & MRC 0134-097  & 1.503 & 0.022 & 13.0 & $<0.56$ & 2 & Double & 01 37 16.63  & -09 31 12.0 &  J013716.67-093111.0  & 16.033 & 0.051 & .. & .. &  \\
J013800-053330 & MRC 0135-058  & 2.313 & 0.018 & 15.2 & $<0.54$ & 3 & Triple  & 01 38 00.80 & -05 33 35.7 &  J013800.70-053335.5  & 14.677 & 0.029 & .. & .. &  \\
  & \\ 
J013933-085023 & MRC 0137-090  & 2.664 & 0.022 & 19.3 & $<0.38$ & 4 & Double & 01 39 33.22 & -08 50 23.2 &  J013933.12-085027.0 & 15.937 & 0.052 & .. & .. &    \\
J014013-095654 & PKS 0137-10  & 4.351 & 0.021 & 33.5 & 0.96 & 1 & Single & 01 40 12.840 & -09 56 57.39 &  J014012.84-095657.7  & 16.327 & 0.066 & .. & .. & AP   \\
J014144-074607 & MRC 0139-080  & 1.649 & 0.018 & 12.9 & 1.01 & 1 & Single & 01 41 43.986 & -07 46 11.14 &  J014143.95-074611.4  & 17.357 & 0.133 & .. & .. &  \\
J014237-074232 & PKS 0140-07 & 2.426 & 0.018 & 17.2 & 0.68 & 1 & Single & 01 42 36.440 & -07 43 03.29 & .. & .. & .. & .. & .. & $\bigstar$ \\ 
J014645-053758 & PKS 0144-05  & 4.351 & 0.02 & 17.5 & 0.54 & 1 & single & 01 46 45.124 & -05 37 58.11 &  J014645.09-053759.2  & 14.654 & 0.03 & 0.63 & Wo00 & A   \\
  & \\
J015050-090126 & PKS 0148-092   & 8.408 & 0.023 & 42.1 & 0.28 & 1 & Single (ext) & 01 50 50.384 & -09 01 29.00 &  J015050.35-090126.9  & 16.348 & 0.06 & .. & .. & A   \\
J015142-103019 & PKS 0149-107 & 5.21 & 0.02 & 29.7 & $<0.30$ & 2 & Double & 01 51 42.37  & -10 30 20.5  &  J015142.29-103019.8  & 15.556 & 0.046 & .. & .. &   \\
J015323-033359 & PKS 0150-03  & 8.399 & 0.021 & 20.0 & 0.76 & 1 & Single & 01 53 23.896 & -03 33 59.31 &  J015323.84-033359.7  & 16.983 & 0.099 & .. & .. &   \\
  &  \\
\multicolumn{16}{l}{{\bf (b) Strongly-scintillating sources with intermediate S/N in the FIRST overlap area} (NSI $>0.90$, 8.0 $\leq$ NormSNR $<12.5$) } \\
J003931-111057 & PMN J0039-1111 & 1.006	& 0.019 & 11.0	& 1.10 & 1	& Single & 00 39 31.591	& -11 11 02.92 &  J003931.58-111102.6 & 13.907 & 0.027 & 0.5526 & SDSS & A \\
J005315-070232 & PMN J0053-0702 & 1.036 & 0.016 & 9.2 & 0.94 & 1 & Single & 00 53 15.689 & -07 02 32.67 & J005315.67-070232.5 & 16.313	& 0.069 & .. & .. &  \\
J010021-074718 & MRC 0057-080 & 0.942 & 0.015 & 9.9 & 0.90 & 1 & Single & 01 00 21.063  & -07 47 18.23	 & J010021.14-074718.4 & 16.771	& 0.099	& .. & .. &  \\
J011049-074142 & PKS 0108-079  & 0.723	& 0.015 & 9.0 & 0.94 & 1 & Single & 01 10 50.018 & -07 41 41.07 & J011050.00-074141.1 & 15.362	& 0.039	& 1.776	& Wr83 & A \\
J011857-071855 & MRC 0116-075 & 1.109 & 0.017 & 10.0 & 1.61 & 1 & Single & 01 18 57.486 & -07 18 55.93 & J011857.53-071855.6  & 16.656	& 0.087	& .. & .. & P \\
& \\
J012102-094426 & PMN J0121-0944 & 0.783 & 0.02 & 8.9 & 1.25 & 1 & Single & 01 21 02.078 & -09 44 28.40	 & J012102.07-094428.6  & 16.144 & 0.063 & .. & ..	&  \\
& \\
\hline
\end{tabular}
{\bf Table columns: } (1) GLEAM name for the source;  
(2)  Alternative name; 
(3)  162\,MHz flux density in Jy, from \cite{Chhetri2018}; 
(4)  Uncertainty in 162\,MHz flux density; 
(5)  Normalised S/N ratio from \cite{Chhetri2018};   
(6)  Normalised scintillation index, or an upper limit for sources where IPS was not detected; 
(7)  Number of separate FIRST components associated with the GLEAM source;
(8) Radio source classification; 
(9)	Right Ascension used for WISE cross-matching (see text for details);
(10) Declination used for WISE cross-matching; 
(11) Name of matched WISE source (if any); 
(12) WISE W1 magnitude; 
(13) Uncertainty in W1; 
(14) Optical spectroscopic redshift (if available); 
(15)  Reference for optical redshift (see \S2.6 of the text for codes); 
(16) Additional notes:
{\rm A : AT20G source \citep{Murphy2010}}, 
{\rm F : Fermi gamma-ray source \citep[3FGL;][]{Acero2015}}, 
{\rm P : Peaked-spectrum source \citep{Callingham2017} }, 
{$\bigstar$ : {\rm See individual notes in Appendix}}.
\end{table}
\end{landscape}

\newpage

\begin {landscape}
\begin{table}
  \caption{Expanded list of strongly-scintillating (NSI $\geq0.90$, NormSNR $\geq8.0$) sources from the Chhetri et al. (2018a) IPS sample, with no restriction on sky area. This table includes some objects from the main IPS-FIRST sample listed in Table \ref{tab:data2} and the IPS-MRC-1Jy sample listed in Table \ref{tab:data4} -- these are indicated by a $\lozenge$ and a $\dagger$ respectively in column 14. }
  \label{tab:data3}
  \begin{tabular}{llrrrlllllllll}
    \hline
    GLEAM name & Alt name & S$_{162}$ & $\pm$ & Norm & NSI & \multicolumn{2}{c}{NVSS position (J2000)} & WISE & W1 & $\pm$ & z  & z$\_$ref &Notes \\
     & & Jy & Jy & SNR & & RA & Dec & match & mag  \\
(1) & (2) & (3) & (4) & (5) & (6) & (7) & (8) & (9) & (10) & (11) & (12) & (13) & (14) \\
\hline
J000905-205630 & PKS 0006-212 & 1.322 & 0.017 &  11.5  &  1.01  &  00 09 06.02  &  -20 56 32.3  &  J000905.98-205632.5  &  16.166  &  0.059  &  0.91  &  Mc96  &  P  \\
J001343-253709  &  PMN J0013-2537  &  1.319  &  0.019  &  10.5  &  0.93  &  00 13 44.17  &  -25 37 12.2  &  J001344.10-253712.4  &  16.646  &  0.088  &  ..  &  ..  &   \\
J001821-111139  &  MRC 0015-114  &  1.932  &  0.020  &  16.2  &  0.95  &  00 18 22.04  &  -11 11 38.2  &  J001821.92-11139.1 &  17.042  &  0.128  &  ..  &  ..  & $\lozenge$ \\
J001931-043543  &  3C 007  &  4.523  &  0.018  &  16.4  &  1.03  &  00 19 31.49  &  -04 35 47.8  &  J001931.48-043548.1  &  16.573  &  0.087  &  ..  &  ..  & $\lozenge$ \\
J002105-184435  &  NVSS J002105-184438  &  0.976  &  0.017  &  11.1  &  0.90  &  00 21 05.96  &  -18 44 38.0  &  J002106.00-184438.4  &  15.819  &  0.047  &  ..  &  ..  &  \\
& \\
J002208-104133  &  PKS 0019-109  &  1.774  &  0.019  &  14.3  &  1.34  &  00 22 08.97  &  -10 41 32.6  &  J002209.00-104132.1  &  17.419  &  0.175  &  ..  &  ..  &  P $\lozenge$  \\
J002223-070230  &  PKS 0019-073  &  3.821  &  0.016  &  23.1  &  1.25  &  00 22 23.04  &  -07 02 35.1  &  ..  & .. &  &  ..  &  ..  &  $\bigstar$ P $\lozenge$ \\
J002546-124724  &  MRC 0023-130B  &  1.388  &  0.021  &  15.4  &  1.14  &  00 25 46.66  &  -12 47 25.2  &  J002546.68-124724.9  &  15.850  &  0.051  &  ..  &  ..  &  \\
J003119-165202  &  PMN  J0031-1651 &  0.718  &  0.015  &  8.5  &  1.08  &  00 31 19.07  &  -16 52 07.7  &  J003119.14-165208.3  &  17.312  &  0.157  &  ..  &  ..  &  P  \\
J003201-085133  &  PKS 0029-091  &  2.125  &  0.020  &  16.6  &  1.27  &  00 32 00.98  &  -08 51 37.6  &  J003200.97-085139.1  &  17.593  &  0.212  &  ..  &  ..  &  P $\lozenge$ \\
& \\
J003242-123339  &  PMN J0032-1233  &  1.132  &  0.019  &  11.6  &  1.16  &  00 32 42.57  &  -12 33 39.2  &  J003242.28-123338.0  &  15.941 & 0.056  &  ..  &  ..  & $\bigstar$ \\
J003829-211957  &  PKS  0035-216  &  1.081  &  0.017  &  13.9  &  1.12  &  00 38 29.93  &  -21 20 04.0  &  J003829.96-212004.0  &  14.286  &  0.027  &  0.338  &  Mc96  &  AP $\dagger$ \\
J003931-111057  &  PMN J0039-1111  &  1.006  &  0.019  &  11.0  &  1.10  &  00 39 31.57  &  -11 11 02.3  &  J003931.58-111102.6  &  13.907  &  0.027  &  0.5526  &  SDSS  &  A \\
J004106-131440  &  TXS 0038-135  &  0.637  &  0.020  &  8.4  &  1.14  &  00 41 06.43  &  -13 14 39.7  &  J004106.42-131438.2  &  16.116  &  0.064  &  ..  &  ..  &   \\
J004258-203649  &  PKS 0040-208  &  1.207  &  0.016  &  15.0  &  0.99  &  00 42 58.38  &  -20 37 13.0  &  J004258.36-203713.2  &  12.321  &  0.023  &  0.6554  &  6dFGS  & $\dagger$  \\
& \\
J004537-273534  &  PMN J0045-2735  &  0.829  &  0.015  &  9.5  &  1.25  &  00 45 37.13  &  -27 35 38.1  &  J004537.04-273538.8  &  16.001  &  0.056  &  ..  &  ..  &  P  \\
J004807-183838  &  MRC 0045-189  &  1.340  &  0.015  &  18.0 &  1.12  &  00 48 07.55  &  -18 38 41.6  &  J004807.64-183840.3  &  16.836  &  0.116  &  ..  &  ..  &  \\
J004839-294718  &  PKS 0046-300  &  1.611  &  0.015  &  16.33  &  1.25  &  00 48 39.49  &  -29 47 18.5  &  ..  & 19.4 & 0.9 &  ..  &  ..  &  $\bigstar$ P  \\
J004954-100613  &  PKS 0047-10  &  2.257  &  0.018  &  27.3  &  1.05  &  00 49 54.09  &  -10 06 14.4  &  J004954.03-100614.5  &  16.183  &  0.061  &  2.236  &  SDSS  &  A $\lozenge$  \\
J005026-120115  &  PKS 0048-12  &  1.950  &  0.020  &  25.5  &  1.05  &  00 50 25.94  &  -12 01 17.0  & J005025.57-120110.4  & 14.814 &  0.033  &  ..  &  ..  &  $\bigstar$ A  \\
 & \\
J005027-175237  &  TXS 0047-181  &  0.668  &  0.016  & 9.5  &  1.17  &  00 50 27.67  &  -17 52 39.6  &  ..  & 18.31 & 0.30 &  ..  &  ..  &  $\bigstar$ \\
J005039-102734  &  MRC 0048-107  &  2.138  &  0.018  &  26.8  &  0.90  &  00 50 39.04  &  -10 27 35.2  &  ..  & 15.67 &  0.05  &  ..  &  ..  & $\bigstar$  $\lozenge$ \\
J005234-093819  &  MRC 0050-099  &  1.472  &  0.018  &  17.2  &  1.19  &  00 52 34.53  &  -09 38 19.2  &  ..  & .. & ..  &  ..  &  ..  & $\lozenge$  \\
J005315-070232  &  PMN J0053-0702  &  1.036  &  0.016  &  9.2  &  0.94  &  00 53 15.65  &  -07 02 33.4  &  J005315.67-070232.5  &  16.313  &  0.069  &  ..  &  ..  &  \\
J005429-235127  &  MRC 0052-241  &  2.818  &  0.017  &  38.8  &  0.92  &  00 54 29.85  &  -23 51 31.2  & .. & .. & .. &  2.86  &  Mc96  & $\dagger$ \\
& \\
J005433-195255  &  PMN  J0054-1952  &  0.726  &  0.016  &  10.7  &  1.06  &  00 54 32.94  &  -19 53 00.7  &  J005432.93-195301.1  &  15.856  &  0.051  &  ..  &  ..  &  A  \\
J005654-201828  &  NVSS J005653-201834  &  0.639  &  0.015  &  8.7  &  1.35  &  00 56 53.97  &  -20 18 34.3  & .. & .. & .. &  ..  &  ..  &   \\
J005930-184204  &  NVSS  J005930-184209  &  0.572  &  0.016  &  9.0  &  1.24  &  00 59 30.25  &  -18 42 09.8  &  J005930.27-184209.6  &  14.527  &  0.03  &  ..  &  ..  &  \\
J010021-074718  &  MRC 0057-080  &  0.942  &  0.015  &  9.9   &  0.90  &  01 00 21.01  &  -07 47 17.8  &  J010021.14-074718.4  &  16.771  &  0.099  &  ..  &  ..  & \\
J010042-050237  &  MRC 0058-053  &  1.910  &  0.016  &  13.2  &  1.34  &  01 00 42.67  &  -05 02 36.4  &  J010042.64-050236.5  &  15.987  &  0.052  &  ..  &  ..  &  P $\lozenge$ \\
& \\
J010152-283118  &  PKS 0059-287  &  0.983  &  0.014  &  10.0  &  1.20  &  01 01 52.38  &  -28 31 19.7  &  J010152.37-283120.5  &  16.020  &  0.052  &  1.6  &  Ka98  &  AP  \\
J010202-261447  &  NVSS J010202-261450  &  0.878  &  0.016  &  11.1  &  1.11  &  01 02 02.71  &  -26 14 50.8  &  J010202.68-261451.3  &  17.427  &  0.156  &  ..  &  ..  &  \\
J010250-140230  &  PMN J0102-1402  &  0.804  &  0.016  &  11.95  &  1.03  &  01 02 51.10  &  -14 02 33.2  &  J010251.06-140233.4  &  16.702  &  0.091  &  ..  &  ..  &  \\
J010316-322550  &  PKS 0100-326  &  3.615  &  0.012  &  24.5 &  0.98  &  01 03 16.77  &  -32 25 51.8  &  J010316.79-322552.3  &  14.342  &  0.028  &  0.256  &  Dr03  &  P  \\
J010335-271506  &  PKS 0101-275  &  3.077  &  0.015  &  36.3  &  0.99  &  01 03 35.49  &  -27 15 11.1  &  J010335.60-271510.3  &  17.589  &  0.178  &  ..  &  ..  &  P $\dagger$ \\
& \\
\hline
  \end{tabular}
 \end{table}
\end{landscape}

\begin {landscape}
\begin{table}
  \begin{tabular}{llrrrlllllllll}
    \hline
    GLEAM name & Alt name & S$_{162}$ & $\pm$ & Norm & NSI & \multicolumn{2}{c}{NVSS position (J2000)} & WISE & W1 & $\pm$ & z  & z$\_$ref & Notes \\
     & & Jy & Jy & SNR & & RA & Dec & match & mag  \\
(1) & (2) & (3) & (4) & (5) & (6) & (7) & (8) & (9) & (10) & (11) & (12) & (13) & (14) \\
\hline
J010452-252052  &  PKS 0102-256  &  3.391  &  0.016  &  44.2  &  1.06  &  01 04 52.32  &  -25 20 53.3  &  J010452.24-252054.5  &  17.130  &  0.132  &  ..  &  ..  & $\dagger$  \\
J010527-182843  &  PKS 0103-187  &  1.028  &  0.016  &  16.0  &  1.00  &  01 05 27.04  &  -18 28 45.7  &  J010527.05-182846.3  &  15.175  &  0.036  &  ..  &  ..  &  A  \\
J010607-114246  &  MRC 0103-119  &  1.511  &  0.020  &  20.3 &  1.18  &  01 06 7.98  &  -11 42 46.6  &  ..  & .. & .. &  ..  &  ..  &  P  \\
J010617-233615  &  PMN J0106-2336  &  0.594  &  0.018  &  9.0  &  1.07  &  01 06 17.42  &  -23 36 16.0  &  J010617.45-233616.9  &  15.111  &  0.036  &  ..  &  ..  &  P  \\
J010703-222257  &  PMN J0107-2223  &  0.725  &  0.015  &  9.7  &  1.06  &  01 07 03.83  &  -22 23 00.2  &  J010703.84-222300.8  &  17.005  &  0.112  &  ..  &  ..  &  P  \\
 & \\
J010837-285124  &  PKS 0106-291  &  5.040  &  0.014  &  50.0  &  1.17  &  01 08 37.88  &  -28 51 27.7  &  J010837.89-285129.4  &  16.916  &  0.102  &  ..  &  ..  &  AP $\dagger$ \\
J010938-144228  &  MRC 0107-149  &  2.864  &  0.020  &  39.1  &  0.93  &  01 09 38.83  &  -14 42 34.9  &  J010938.86-144233.9  &  16.783  &  0.100  &  ..  &  ..  &   \\
J011049-074142  &  PKS 0108-079  &  0.723  &  0.015  &  9.0  &  0.94  &  01 10 50.01  &  -07 41 41.6  &  J011050.00-074141.1  &  15.362  &  0.039  &  1.776  &  Wr83  &  A \\
J011156-302623  &  TXS 0109-307  &  0.990  &  0.013  &  9.2  &  1.24  &  01 11 55.97  &  -30 26 26.2  &   J011156.00-302622.2  & 16.019 &  0.055 & .. & ..  &  $\bigstar$ P  \\ 
J011312-101419  &  PKS  0110-105  &  1.565  &  0.019  &  20.2 &  1.46  &  01 13 12.08  &  -10 14 21.8  &  J011312.09-101421.2  &  15.285  &  0.037  &  1.3210  &  SDSS  &  A  \\
 & \\
J011651-205202  & PKS 0114-21  &  14.387  &  0.020  &  176.6  &  0.90  &  01 16 51.43  &  -20 52 06.6  &  J011651.48-205207.1  &  16.363  &  0.069  &  1.41  &  Mc96  &  AP $\dagger$ \\
J011657-141105  &  TXS 0114-144  &  0.638  &  0.020  &  8.3  &  1.06  &  01 16 57.88  &  -14 11 10.5  &  J011657.86-141110.2  &  16.178  &  0.056  &  ..  &  ..  &    \\
J011738-150750  &  PKS 0115-153  &  0.921  &  0.021  &  12.6  &  1.15  &  01 17 38.93  &  -15 07 54.6  &  J011738.93-150754.8  &  16.208  &  0.059  &  ..  &  ..  &   \\
J011806-135542  &  TXS 0115-141  &  0.806  &  0.021  &  9.2  &  1.43  &  01 18 06.29  &  -13 55 48.0  &  J011806.36-135547.3  &  17.525  &  0.148  &  ..  &  ..  &  P  \\
J011815-012037  &  PKS 0115-01  &  5.357  &  0.020  &  16.4  &  1.20  &  01 18 15.36  &  -01 20 30.3  &  J011815.39-012030.2  &  15.557  &  0.047  &  1.162  &  Ho03  &  AP $\lozenge$ \\
  & \\
J011823-132620  &  MRC 0115-137  &  2.194  &  0.020  &  26.6  &  1.02  &  01 18 24.17  &  -13 26 23.9  &  J011824.19-132624.0  &  16.273  &  0.059  &  ..  &  ..  &    \\
J011857-071855  &  MRC  0116-075  &  1.109  &  0.017  &  10.0  &  1.61  &  01 18 57.49  &  -07 18 56.7  &  J011857.53-071855.6  &  16.656  &  0.087  &  ..  &  ..  &  P \\
J011915-162313  &  TXS 0116-166B  &  1.357  &  0.019  &  16.4  &  0.93  &  01 19 15.27  &  -16 23 19.1  &  J011915.31-162319.2  &  17.229  &  0.126  &  ..  &  ..  &    \\
J011930-113302  &  MRC 0117-118  &  1.266  &  0.019  &  15.5  &  1.13  &  01 19 30.52  &  -11 33 07.6  &  ..  & .. &  &  ..  &  ..  &  P  \\
J012102-094426  &  PMN J0121-0944  &  0.783  &  0.020  &  8.9  &  1.25  &  01 21 2.11  &  -09 44 29.7  &  J012102.07-094428.6  &  16.144  &  0.063  &  ..  &  ..  &  \\
 & \\
J012114-182719  &  PKS 0118-187  &  1.074  &  0.017  &  14.2  &  1.16  &  01 21 14.51  &  -18 27 22.5  &  J012114.48-182724.6  &  17.739  &  0.202  &  ..  &  ..  &  P  \\
J013056-150232  &  PKS 0128-152  &  2.553  &  0.019  &  30.7  &  0.91  &  01 30 56.73  &  -15 02 36.0  &  J013056.69-150235.6  &  16.378  &  0.073  &  ..  &  ..  &   \\
J013212-065232  &  PKS 0129-07  &  5.309  &  0.020  &  36.7  &  1.19  &  01 32 12.20  &  -06 52 36.8  &  J013212.24-065236.0  &  15.951  &  0.051  &  ..  &  ..  &  AP $\lozenge$ \\
J013243-165444  &  PKS 0130-17  &  1.453  &  0.018  &  14.0  &  0.92  &  01 32 43.41  &  -16 54 47.7  &  J013243.48-165448.5  &  12.309  &  0.025  &  1.02  &  Wr83  &  AFP  \\
J013919-124741  &  TXS 0136-130  &  1.029  &  0.019  &  8.3  &  1.16  &  01 39 19.91  &  -12 47 42.5  &  ..  & 18.29 & 0.32 &  ..  &  ..  &  $\bigstar$ P  \\
 & \\
J014013-095654  &  PKS 0137-10  &  4.351  &  0.021  &  33.5  &  0.96  &  01 40 12.82  &  -09 56 57.1  &  J014012.84-095657.7  &  16.327  &  0.066  &  ..  &  ..  &  AP $\lozenge$ \\
J014045-291744  &  MRC 0138-295  &  1.554  &  0.016  &  8.3  &  1.10  &  01 40 45.82  &  -29 17 51.2  &  J014045.87-291750.4  &  16.300  &  0.063  &  ..  &  ..  &    \\
J014144-074607  &  MRC 0139-080  &  1.649  &  0.018  &  12.9  &  1.01  &  01 41 44.06  &  -07 46 11.8  &  J014143.95-074611.4  &  17.357  &  0.133  &  ..  &  ..  & $\lozenge$  \\
J015208-192323  &  MRC 0149-196  &  1.391  &  0.015  &  10.8  &  1.12  &  01 52 08.73  &  -19 23 25.4  &  J015208.71-192325.2  &  16.809  &  0.092  &  ..  &  ..  &    \\
J015455-204023  &  MRC 0152-209  &  2.471  &  0.014  &  15.4  &  1.14  &  01 54 55.73  &  -20 40 26.9  &  .. 
& 16.259 & 0.036  &  1.9212  &  Mc91  & $\bigstar$  $\dagger$ \\
 &  \\
\hline
  \end{tabular}

{\bf Table columns: } (1) GLEAM name for the source;  
(2)  Alternative name; 
(3)  162\,MHz flux density in Jy, from \cite{Chhetri2018}; 
(4)  Uncertainty in 162\,MHz flux density; 
(5)  Normalised S/N ratio from \cite{Chhetri2018};   
(6)  Normalised scintillation index, or an upper limit for sources where IPS was not detected; 
(7)	NVSS Right Ascension;  
(8)	NVSS Declination; 
(9) Name of matched WISE source (if any);  
(10) WISE W1 magnitude; 
(11) Uncertainty in W1; 
(12) Optical spectroscopic redshift (if available); 
(13) Reference for optical redshift (see \S2.6 of the text for codes); 
(14) Additional notes: 
{\rm A : AT20G source \citep{Murphy2010}}, 
{\rm F : Fermi gamma-ray source \citep[3FGL;][]{Acero2015}}, 
{\rm P : Peaked-spectrum source \citep{Callingham2017} }, 
{$\bigstar$ : {\rm See individual notes in Appendix}},
$\lozenge$ {\rm : Source also included in Table \ref{tab:data2} },
$\dagger$ {\rm : Source also included in Table \ref{tab:data4}. }
\end{table}
\end{landscape}

\newpage

\begin {landscape}
\begin{table}
  \caption{The IPS-MRC-1Jy sample of sources from the Chhetri et al. (2018a) IPS sample that also appear in the MRC-1Jy sample \citep{McCarthy1996,Kapahi1998b}. }
  \label{tab:data4}
  \begin{tabular}{llrrrrllllllllllll}
    \hline
GLEAM name & MRC name & S162 & $\pm$ & Norm & NSI & \multicolumn{2}{c}{Radio position (J2000)} &  T & WISE & W1 & $\pm$ & z & z\_ref & Notes  \\
                        &                    &  Jy     &  Jy       & SNR  &       &  RA          & Dec         &     & match  & mag &        &    &           &       \\
  (1) & (2) & (3) & (4) & (5) & (6) & (7) & (8) & (9) & (10) & (11) & (12) & (13) & (14) & (15) \\
\hline
J000347-232938 & MRC 0001-237 & 3.040 & 0.021 & 20.3 & $<0.17$ & 00 03 47.89 & -23 29 42.1 & gal & J000347.92-232941.6 & 14.398 & 0.030 & 0.315 & Mc96 &  \\
J000402-230657 & MRC 0001-233 & 2.023 & 0.020 & 13.9 & $<0.44$ & 00 04 02.55 & -23 06 57.4 & gal &  J000402.48-230657.0 & 13.507 & 0.025 & 0.097 & Mc96 &   \\ 
J000957-282930 & MRC 0007-287 & 2.212 & 0.019 & 14.1 & $<0.37$ & 00 09 58.67 & -28 29 30.1 & gal & J000958.73-282930.0 & (12.821) & 0.023 & .. & .. & $\bigstar$  \\
J001757-223800 & MRC 0015-229 & 2.385 & 0.017 & 22.7 & $<0.27$ & 00 17 58.23 & -22 38 03.4 & gal & J001758.16-223803.2 & 16.870 & 0.109 & 2.01 & Mc96 &   \\
J002021-202846 & MRC 0017-207  & 1.946 & 0.019 & 18.8 & $<0.32$ & 00 20 21.49  &  -20 28 51.4   &  QSO  &  J002021.48-202851.0 & 14.750 & 0.030 & 0.545 & Ba99 &   \\
& \\
J002026-201427 & MRC 0017-205 & 2.714 & 0.020 & 26.6 & $<0.06$ & 00 20 33.18 & -20 16 10.5 & gal & J002033.23-201611.2 & 14.392 & 0.028 & 0.197 & Mc96 &   \\
J002308-250232 & MRC 0020-253 & 8.972 & 0.020 & 81.7 & 0.10 & 00 23 08.90 & -25 02 29.1 & gal &  J002308.86-250229.6 & 15.564 & 0.066 & 0.35 & Mc96 &  \\
J002430-292847 & MRC 0022-297 & 15.271 & 0.018 & 104.7 & 0.13 & 00 24 30.14   &  -29 28 54.6   &  QSO  &  J002430.15-292854.3 & 13.392 & 0.026 & 0.4065 & Ba99 & A  \\
J002448-204259 & MRC 0022-209 & 1.567 & 0.018 & 16.2 & $<0.16$ & 00 24 47.81 & -20 42 39.9 & gal &  J002447.76-204239.7 & 13.422 & 0.025 & 0.054 & Mc96 &  \\
J002549-260211 & MRC 0023-263 & 18.786 & 0.020 & 153.7 & 0.78 & 00 25 49.29 & -26 02 12.9 & gal & J002549.22-260212.3 & 14.559 & 0.031 & 0.322 & Mc96 & AP  \\
 & \\
J002613-200455 & MRC 0023-203 & 5.736 & 0.017 & 61.2 & 0.23 & 00 26 14.11 & -20 04 56.4 & gal &  J002614.09-200456.1 & 15.636 & 0.047 & 0.845 & Mc96 & A \\
J002729-273113 & MRC 0025-277 & 2.676 & 0.018 & 23.5 & 0.32 & 00 27 29.32 & -27 31 12.1 & gal & J002729.11-273113.4 & 17.718 & 0.199 & .. & .. &  \\
J002752-200738 & MRC 0025-204 & 1.566 & 0.017 & 17.4 & 0.40 & 00 27 52.80 & -20 07 38.4 & gal & J002752.86-200738.7 & 15.472 & 0.042 & .. & .. &   \\
J003102-220701 & MRC 0028-223 & 1.745 & 0.015 & 21.0 & $<0.23$ & 00 31 02.46 & -22 07 08.3 & gal & J003102.44-220708.3 & 14.082 & 0.027 & 0.205 & Mc96  & \\
J003150-265223 & MRC 0029-271 & 2.282 & 0.017 & 21.8 & 0.57 & 00 31 50.4   &  -26 52 25   &  QSO  &  J003150.49-265224.6 & 13.869 & 0.025 & 0.333 & Ba99 &  \\
 &&& \\
J003203-225759 & MRC 0029-232 & 2.554 & 0.017 & 29.2 & 0.27 & 00 32 03.60 & -22 58 08.6 & gal &  J003203.30-225804.1 & 16.772 & 0.090 & .. & .. &  & \\
J003221-240505 & MRC 0029-243 & 3.973 & 0.018 & 49.2 & 0.39 & 00 32 21.33 & -24 05 09.1 & gal &  J003221.32-240507.8 & 16.118 & 0.057 & 1.29 & Mc96 &  \\
J003244-214414 & MRC 0030-220  & 1.785 & 0.016 & 21.3 & 0.29 & 00 32 44.7   &  -21 44 22   &  QSO &  J003244.71-214421.9 & 15.612 & 0.049 & 0.806 & Ba99 &  \\ 
J003246-293107 & MRC 0030-297 & 2.632 & 0.016 & 21.2 & 0.71 & 00 32 46.26 & -29 31 10.0 & gal &  J003246.22-293110.7 & 16.529 & 0.077 & .. & .. & P  \\
J003323-214154 & MRC 0030-219 & 1.686 & 0.016 & 20.1 & 0.86 & 00 33 23.88 & -21 42 00.4 & gal & J003323.85-214201.7 & 17.538 & 0.191 & 2.168 & Mc96 &   \\ 
&&& \\
J003508-200354 & MRC 0032-203 & 12.487 & 0.017 & 160.8 & 0.47 & 00 35 09.21 & -20 03 55.3 & gal & J003508.79-200359.4 & 14.096 & 0.027 & 0.516 & Mc96 & A \\
J003722-230825 & MRC 0034-234 & 3.563 & 0.017 & 43.9 & 0.14 & 00 37 22.14 & -23 08 30.2 & gal &  J003722.27-230831.0 & 16.161 & 0.060 & .. & .. &  \\
J003824-225256 & MRC 0035-231 & 2.327 & 0.016 & 29.7 & 0.46 & 00 38 24.82 & -22 53 02.6 & gal & J003824.93-225302.5 & 15.470 & 0.043 & 0.685 & Mc96 &   \\
J003829-211957 & MRC 0036-216 & 1.081 & 0.017 & 13.9 & 1.12 & 00 38 30.03 & -21 20 04.7 & gal &  J003829.96-212004.0 & 14.286 & 0.027 & 0.338 & Mc96 & AP \\
J003956-253425 & MRC 0037-258 & 2.591 & 0.016 & 31.2 & $<0.22$ & 00 39 56.40 & -25 34 30.0 & gal & J003956.45-253431.0 & 15.902 & 0.052 & 1.1 & Mc96 &  \\
 &&& \\
J004048-204329 & MRC 0038-209 & 1.710 & 0.017 & 21.2 & 0.39 & 00 40 48.13 & -20 43 39.9 & gal &  J004048.20-204339.6 & 12.775 & 0.024 & 0.0906 & Mc96 & A \\ 
J004112-290744 & MRC 0038-294 & 1.967 & 0.016 & 21.1 & $<0.29$ & 00 41 12.14 & -29 07 46.9 & gal & J004111.62-290748.3 & 16.460 & 0.064 & .. & .. &  \\
J004258-203649 & MRC 0040-208  & 1.207 & 0.016 & 15.0 & 0.99 & 00 42 58.4   &  -20 37 14   &  QSO  & J004258.36-203713.2 & 12.321 & 0.023 & 0.6554 & Ba99 &  \\
J004411-221219 & MRC 0041-224 & 3.935 & 0.016 & 53.8 & 0.31 & 00 44 11.98 & -22 12 18.3 & gal & .. & .. & .. & .. & .. &  \\
J004503-243417 & MRC 0042-248 & 2.644 & 0.017 & 33.5 & $<0.18$ & 00 45 03.12 & -24 34 23.6 & gal &  J004502.95-243423.8 & 14.635 & 0.031 & .. & .. &  & \\
&&& \\
J004733-251710 & MRC 0045-255 & 6.577 & 0.020 & 66.7 & 0.07 & 00 47 27.41 & -25 13 37.9 & gal &  J004733.14-251717.7 & 5.757 & 0.034 & 0.001 & RC3 & AF  \\
J005053-230650 & MRC 0048-233 & 1.854 & 0.016 & 25.5 & $<0.20$ & 00 50 53.04 & -23 06 53.8 & gal &  J005052.87-230654.1 & 13.287 & 0.046 & 0.111 & Mc96 &   \\
J005242-215540 & MRC 0050-222 & 2.454 & 0.016 & 34.1 & 0.85 & 00 52 42.79 & -21 55 45.2 & gal &  J005242.84-215546.7 & 15.635 & 0.045 & 0.654 & Mc96 &  \\ 
J005429-235127 & MRC 0052-241 & 2.818 & 0.017 & 38.8 & 0.92 & 00 54 29.82 & -23 51 29.7 & gal &  .. & .. & .. & 2.86 & Mc96 &   \\
J005742-253313 & MRC 0055-258 & 1.049 & 0.017 & 12.8 & $<0.30$ & 00 57 42.64 & -25 33 14.4 & gal &  J005742.75-253315.6 & 15.054 & 0.038 & .. & .. &   \\
 &&& \\
\hline
  \end{tabular}
 \end{table}
\end{landscape}

\newpage

\begin{landscape}
\begin{table}
  \begin{tabular}{llrrrrllllllllllll}
    \hline
GLEAM name & MRC name & S162 & $\pm$ & Norm & NSI & \multicolumn{2}{c}{Radio position (J2000)} &  T & WISE & W1 & $\pm$ & z & z\_ref & Notes  \\
                        &                    &  Jy     &  Jy       & SNR  &       &  RA          & Dec         &     & match  & mag &        &    &           &       \\
  (1) & (2) & (3) & (4) & (5) & (6) & (7) & (8) & (9) & (10) & (11) & (12) & (13) & (14) & (15) \\
\hline
& \\
J005756-252233 & MRC 0055-256 & 1.301 & 0.017 & 16.6 & $<0.35$ & 00 57 57.05 & -25 22 36.7 & gal &  J005756.98-252237.4 & 14.135 & 0.027 & 0.1985 & Mc96 &  \\ 
J005827-240101 & MRC 0056-242 & 3.676 & 0.016 & 53.0 & $<0.11$ & 00 58 27.79 & -24 01 05.4 & gal & J005827.81-240104.8 & 14.727 & 0.030 & .. & .. &  \\
J010041-223955 & MRC 0058-229 & 2.605 & 0.016 & 36.0 & $<0.15$ & 01 00 42.0   &  -22 39 58   &  QSO  & J010042.10-223959.8 & 14.165 & 0.027 & 0.706 & Ba99 &   \\
J010241-215227 & MRC 0100-221 & 9.191 & 0.018 & 114.4 & 0.04 & 01 02 41.69 & -21 52 53.8 & gal & J010241.76-215254.2 & 11.451 & 0.022 & 0.058 & Mc96 &  \\ 
J010244-273124 & MRC 0100-277 & 5.929 & 0.015 & 68.4 & 0.20 & 01 02 43.92 & -27 31 24.8 & gal & J010244.00-273126.0 & 15.540 & 0.041 & .. & .. &   \\
  &&& \\
J010335-271506 & MRC 0101-275 & 3.077 & 0.015 & 36.3 & 0.99 & 01 03 35.48 & -27 15 09.2 & gal & J010335.60-271510.3 & 17.589 & 0.178 & .. & .. & P  \\
J010452-252052 & MRC 0102-256 & 3.391 & 0.016 & 44.2 & 1.06 & 01 04 52.17 & -25 20 53.1 & gal & J010452.24-252054.5 & 17.130 & 0.132 & .. & .. &   \\
J010612-240634 & MRC 0103-243 & 2.076 & 0.017 & 27.5 & 0.22 & 01 06 12.61 & -24 06 38.4 & gal &  J010612.70-240635.5 & 17.014 & 0.104 & .. & .. &   \\
J010837-285124 & MRC 0106-291 & 5.040 & 0.014 & 50.0 & 1.17 & 01 08 37.98 & -28 51 28.6 & gal & J010837.89-285129.4 & 16.916 & 0.102 & .. & .. & AP  \\
J010902-230728 & MRC 0106-233  & 1.937 & 0.015 & 26.9 & 0.63 & 01 09 03.0   &  -23 07 29   &  QSO  & J010903.02-230729.7 & 14.920 & 0.035 & 0.818 & Ba99 &  \\ 
 &&& \\
J011228-221045 & MRC 0110-224 & 2.390 & 0.015 & 32.6 & 0.55 & 01 12 28.92 & -22 10 49.1 & gal &J011228.95-221051.0 & 15.846 & 0.052 & .. & .. &   \\
J011342-252358 & MRC 0111-256  & 1.606 & 0.016 & 19.9 & $<0.29$ & 01 13 42.6   &  -25 23 59   &  QSO  & J011342.59-252359.0 & 15.843 & 0.047 & 1.05 & Ba99 & A \\ 
J011442-204239 & MRC 0112-209 & 4.638 & 0.017 & 59.9 & 0.28 & 01 14 42.36 & -20 42 49.7 & gal & J011442.22-204249.0 & 16.307 & 0.069 & .. & .. &   \\
J011516-213834 & MRC 0112-219 & 1.786 & 0.016 & 21.6 & $<0.19$ & 01 15 16.72 & -21 38 37.5 & gal &  J011516.99-213843.1 & 15.451 & 0.041 & .. & .. &  \\ 
J011539-281707 & MRC 0113-285 & 2.950 & 0.015 & 30.6 & $<0.27$ & 01 15 39.52 & -28 17 07.2 & gal &  J011539.54-281709.5 & 15.788 & 0.044 & .. & .. & A  \\
 &&& \\
J011544-241719 & MRC 0113-245 & 2.499 & 0.016 & 29.7 & 0.34 & 01 15 44.61 & -24 17 14.1 & gal &  J011544.66-241715.0 & 15.331 & 0.039 & .. & .. & \\
J011651-205202 & MRC 0114-211 & 14.387 & 0.020 & 176.6 & 0.90 & 01 16 51.44 & -20 52 07.5 & gal  & J011651.48-205207.1 & 16.363 & 0.069 & 1.41: & Mc96 & AP \\ 
J011815-255148 & MRC 0115-261 & 4.329 & 0.016 & 48.6 & $<0.15$ & 01 18 15.66 & -25 51 49.4 & gal  & J011815.76-255149.6 & 14.629 & 0.029 & 0.268 & Mc96 & A  \\
J012031-270125 & MRC 0118-272  & 1.866 & 0.016 & 20.1 & 0.51 & 01 20 31.6   &  -27 01 25   &  QSO   & J012031.66-270124.6 & 11.316 & 0.022 & ... & .. & AF  \\ 
J012331-291629 & MRC 0121-295 & 2.725 & 0.015 & 21.4 & $<0.28$ & 01 23 31.50 & -29 16 29.6 & gal  & J012331.51-291630.7 & 15.421 & 0.041 & .. & .. &   \\
 &&& \\
J012445-251714 & MRC 0122-255 & 6.909 & 0.016 & 77.5 & 0.39 & 01 24 46.28 & -25 17 13.6 & gal  & J012446.45-251716.9 & 9.852 & 0.024 & .. & .. &   \\
J012614-222227 & MRC 0123-226  & 2.567 & 0.016 & 29.3 & 0.29 & 01 26 15.0   &  -22 22 34   &  QSO  & J012614.99-222233.6 & 13.877 & 0.026 & 0.72 & Ba99 & AF \\
J012730-195607 & MRC 0125-201 & 1.878 & 0.016 & 20.4 & 0.63 & 01 27 30.66 & -19 56 10.4 & gal & .. & .. & .. & .. & .. & .. &  \\
J012808-212207 & MRC 0125-216 & 1.975 & 0.014 & 22.1 & $<0.17$ & 01 28 08.73 & -21 22 10.9 & gal &  J012808.59-212211.2 & 14.179 & 0.029 & 0.34 & Mc96 &  \\
J013009-272440 & MRC 0127-276 & 1.618 & 0.017 & 14.7 & $<0.53$ & 01 30 09.29 & -27 24 41.4 & gal & J013008.96-272441.9 & 15.504 & 0.046 & 0.318 & Mc96 &  \\
 &&& \\
J013027-260956 & MRC 0128-264 & 11.522 & 0.017 & 102.2 & 0.35 & 01 30 27.81 & -26 09 55.8 & gal &  .. & .. & .. & 2.348 & Be99 &   \\
J013525-262333 & MRC 0133-266  & 2.268 & 0.017 & 19.4 & $<0.16$ & 01 35 25.8   &  -26 23 32   &  QSO & J013525.77-262337.7 & 15.187 & 0.035 & 1.53 & Ba99 &   \\
J013600-272742 & MRC 0133-277 & 2.266 & 0.016 & 17.3 & $<0.03$ & 01 36 00.45 & -27 27 44.3 & gal & J013600.50-272744.3 & 16.335 & 0.071 & .. & .. &   \\
J013737-243048 & MRC 0135-247  & 3.760 & 0.016 & 35.9 & 0.27 & 01 37 38.3   &  -24 30 54   &  QSO  & J013738.34-243053.9 & 12.920 & 0.024 & 0.8375 & Ba99 & AF  \\
J013857-225444 & MRC 0136-231  & 2.468 & 0.016 & 23.6 & $<0.28$ & 01 38 57.4   &  -22 54 47   &  QSO  &  J013857.46-225447.3 & 14.892 & 0.032 & 1.893 & Ba99 & A  \\
 &&& \\
J013952-260648 & MRC 0137-263 & 2.772 & 0.016 & 20.7 & $<0.29$ & 01 39 52.60 & -26 07 06.3 & gal &  J013952.63-260710.1 & 16.688 & 0.074 & .. & .. &   \\
J014112-213825 & MRC 0138-218 & 1.662 & 0.015 & 15.0 & $<0.39$ & 01 41 12.29 & -21 38 32.7 & gal & J014112.18-213841.6 & 15.840 & 0.054 & .. & .. &  \\
J014127-270606 & MRC 0139-273 & 8.202 & 0.017 & 57.5 & 0.65 & 01 41 27.08 & -27 06 11.4 & gal &  J014127.03-270611.6 & 16.187 & 0.056 & 1.44 & Mc96 & A  \\
J014241-253033 & MRC 0140-257 & 2.148 & 0.016 & 14.7 & 0.58 & 01 42 41.16 & -25 30 34.1 & gal &  J014241.25-253036.5 & 17.615 & 0.151 & 2.64 & Mc96 &  \\
J014521-242333 & MRC 0143-246 & 3.123 & 0.018 & 24.9 & 0.33 & 01 45 21.30 & -24 23 35.0 & gal &  J014521.34-242331.0 & 15.314 & 0.039 & 0.716 & Mc96 &   \\
& \\
\hline
  \end{tabular}
 \end{table}
\end{landscape}

\newpage

\begin{landscape}
\begin{table}
  \begin{tabular}{llrrrrllllllllllll}
    \hline
GLEAM name & MRC name & S162 & $\pm$ & Norm & NSI & \multicolumn{2}{c}{Radio position (J2000)} &  T & WISE & W1 & $\pm$ & z & z\_ref & Notes  \\
                        &                    &  Jy     &  Jy       & SNR  &       &  RA          & Dec         &     & match  & mag &        &    &           &       \\
  (1) & (2) & (3) & (4) & (5) & (6) & (7) & (8) & (9) & (10) & (11) & (12) & (13) & (14) & (15) \\
\hline
& \\
J014709-223234 & MRC 0144-227 & 1.977 & 0.015 & 16.8 & 0.81 & 01 47 09.26 & -22 32 42.0 & gal & J014709.23-223241.0 & 15.733 & 0.045 & 0.6 & Mc96 &   \\
J014913-221129 & MRC 0146-224 & 2.326 & 0.016 & 16.3 & $<0.26$ & 01 49 13.94 & -22 11 37.8 & gal &  J014913.99-221137.0 & 14.748 & 0.034 & 0.36 & Mc96 &  \\ 
J015035-293158 & MRC 0148-297 & 11.405 & 0.019 & 43.6 & $<0.11$ & 01 50 35.86 & -29 31 56.0 & gal &  J015035.94-293155.3 & 14.636 & 0.029 & 0.41 & Mc96 &   \\
J015223-271855 & MRC 0150-275 & 3.659 & 0.017 & 18.0 & 0.44 & 01 52 23.49 & -27 18 55.7 & gal &  J015223.54-271856.6 & 15.788 & 0.045 & .. & .. &   \\
J015455-204023 & MRC 0152-209 & 2.471 & 0.014 & 15.4 & 1.14 & 01 54 55.72 & -20 40 26.8 & gal & (uncataloged)  & 16.259 & 0.036 & 1.9212 & Mc91 & $\bigstar$  \\
 & \\ 
J015753-210216 & MRC 0155-212 & 3.865 & 0.014 & 20.7 & $<0.31$ & 01 57 53.26 & -21 02 19.3 & gal & J015753.40-210217.0 & 13.101 & 0.023 & 0.159 & Mc96 &   \\
J015814-222036 & MRC 0155-225 & 2.635 & 0.015 & 13.0 & $<0.43$ & 01 58 14.45 & -22 20 42.1 & gal &  J015814.58-222044.7 & 15.680 & 0.045 & .. & .. &   \\
J015833-245928 & MRC 0156-252 & 3.018 & 0.016 & 13.6 & 0.67 & 01 58 33.45 & -24 59 30.2 & gal &  J015833.45-245931.9 & 15.401 & 0.038 & 2.09 & Mc96 &   \\
J234324-214129 & MRC 2340-219 & 4.525 & 0.012 & 14.9 & $<0.18$ & 23 43 24.21 & -21 41 33.8 & gal &  J234324.28-214135.3 & 15.097 & 0.036 & 0.766 & Mc96 &   \\ 
J234545-240232 & MRC 2343-243 & 3.663 & 0.014 & 13.3 & $<0.57$ & 23 45 45.17 & -24 02 31.2 & gal &  J234545.18-240231.0 & 14.044 & 0.027 & 0.6 & Mc96 & A  \\
 &&& \\
J235050-245702 & MRC 2348-252  & 9.403 & 0.015 & 39.3 & $<0.19$ & 23 50 49.7   &  -24 57 03   &  QSO  & J235049.80-245703.5 & 14.159 & 0.029 & 1.39 & Ba99 &   \\
J235128-231708 & MRC 2348-235 & 2.826 & 0.014 & 12.6 & $<0.39$ & 23 51 28.27 & -23 17 05.8 & gal &  J235128.27-231706.3 & 16.041 & 0.057 & 0.952 & Mc96 &   \\
J235352-231126 & MRC 2351-234 & 2.913 & 0.014 & 14.9 & $<0.45 $& 23 53 52.25 & -23 11 29.0 & gal &  J235352.35-231127.9 & 15.275 & 0.038 & 1.03 & Mc96 &  \\ 
J235410-215647 & MRC 2351-222 & 2.656 & 0.013 & 14.0 & $<0.43$ & 23 54 10.55 & -21 56 50.7 & gal &  J235410.61-215651.7 & 15.952 & 0.061 & .. & .. &  \\
J235735-211319 & MRC 2355-214 & 3.321 & 0.013 & 23.4 & 0.73 & 23 57 35.27 & -21 13 27.6 & gal & J235735.42-211322.5 & 16.398 & 0.072 & 1.41 & Mc96 & P  \\
& \\
\hline
  \end{tabular}

{\bf Table columns: } (1) GLEAM name for the source;  
(2)  Alternative name; 
(3)  162\,MHz flux density in Jy, from \cite{Chhetri2018}; 
(4)  Uncertainty in 162\,MHz flux density; 
(5)  Normalised S/N ratio from \cite{Chhetri2018};   
(6)  Normalised scintillation index, or an upper limit for sources where IPS was not detected; 
(7,8) Radio position from the MRC-1Jy catalogue;  
(9)   Type (galaxy or QSO);
(10)  Name of matched WISE source (if any);
(11)  WISE W1 magnitude; 
(12)  Uncertainty in W1; 
(13)  Optical spectroscopic redshift (if available); 
(14)  Reference for optical redshift (see \S2.6 of the text for codes); 
(15)  Additional notes:  
{\rm A : AT20G source \citep{Murphy2010}}, 
{\rm F : Fermi gamma-ray source \citep[3FGL;][]{Acero2015}}, 
{\rm P : Peaked-spectrum source \citep{Callingham2017} }, 
{$\bigstar$ : {\rm See individual notes in Appendix}}.
\end{table}
\end{landscape}


\appendix

\section{Notes on individual sources}

\noindent
{\bf GLEAM J000958.73-282930.0 } (MRC 0007-287) \\
\cite{McCarthy1996} note that this source is occulted by a bright star. As a result, the catalogued W1 magnitude is unreliable and no redshift has yet been measured for the host galaxy. \\

\noindent 
{\bf GLEAM J001356-091952} (PKS\,0011-096) \\
An SDSS spectrum shows strong, narrow optical emission lines, including high-excitation lines of [Ne III] and [Ne V]. \\

\noindent
{\bf GLEAM J002223-070230} (PKS\,0019-073) \\
The host of this radio source remains unidentified. The 
closest WISE source is 10\,arcsec from the FIRST radio position, and is unlikely to be associated. \\

\noindent
{\bf GLEAM J003242-123339} (PMN\,J0032-1233) \\
Based on a visual inspection of the WISE images, we identify the WISE object WISE J003242.28-123338.0 as the host of this compact radio source. Although the WISE object is offset by 4.3 arcsec from the NVSS radio position (which itself has a 1.5 arcsec uncertainty), its W1-W2 colour of 0.84 mag.\ implies that it is a powerful AGN and so is unlikely to be a chance alignment of a foreground galaxy. \\

\noindent
{\bf GLEAM J003251-101801} (MRC\,0030-105) \\
This is a compact (angular separation $\sim10$\,arcsec) double source in FIRST, with a well-defined radio centroid. The closest WISE source is 6.6\,arcsec from the radio centroid and appears to be unrelated. The host of this radio source therefore remains unidentified. \\

\noindent 
{\bf GLEAM J003931-111057} (PMN\,J0039-1111) \\
An SDSS spectrum shows strong optical emission lines, including high-excitation lines of [Ne III], along with a strong, broad Mg II emission line. \\ 

\noindent
{\bf GLEAM J004049-094832} (MRC\,0038-100) \\
There is no catalogued WISE source within 15-20 arcsec of the radio centroid of this double source, and the host of this source remains unidentified. \\

\noindent
{\bf GLEAM J004246-061325} (PKS\,0040-06) \\
This GLEAM source corresponds to the northern hotspot of a wide double with a total angular extent of the radio source of at least 6\,arcmin.The radio position listed by \cite{vanVelzen2015} for this source (J2000: 00 42 45.77 -06 12 41.3) also corresponds to the northern hotspot rather than the whole source. The correct optical identification (with a 17th magnitude galaxy) was first made by \cite{Bolton1971}. The position listed in columns 9 and 10 of Table \ref{tab:data2} is that of the optical galaxy, rather than the FIRST radio centroid. The SDSS spectrum shows narrow optical emission lines superimposed on a stellar continuum. At the SDSS redshift of z=0.1243, the projected linear size of the source is over 800\,kpc. \\

\noindent
{\bf GLEAM J004839-2947189} (PKS\,0046-300) \\
A faint uncatalogued source is visible in the WISE W1 image, at an offset of 2.5\,arcsec from the NVSS radio position. One of us (THJ) measured a magnitude of W1 = $19.4\pm0.9$ mag. for this object, which we tentatively identify as the host of the radio source. This WISE source was not detected in the W2-W4 bands. \\

\begin{figure}
\includegraphics[width=\columnwidth]{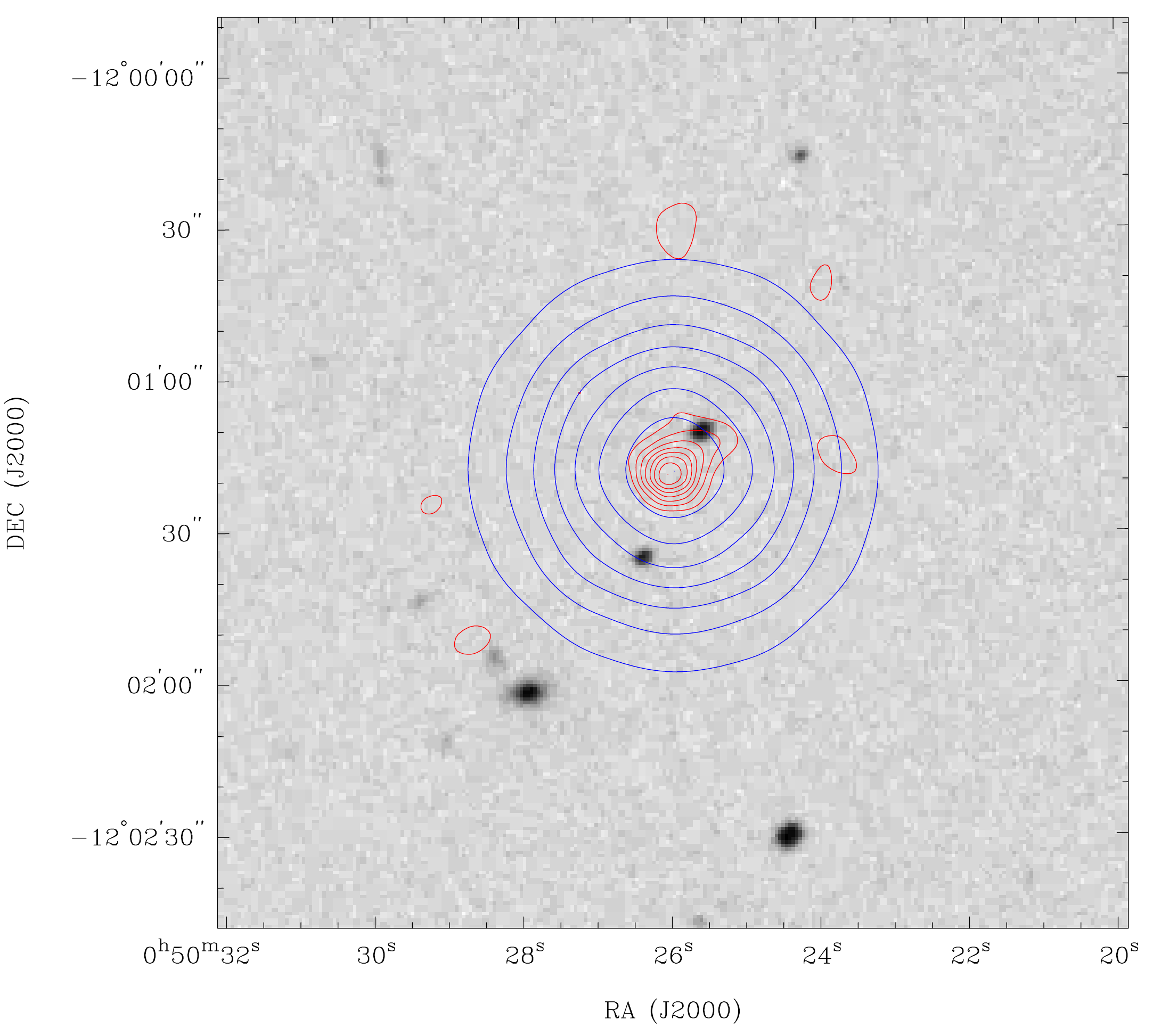}
  \caption{Radio contours from the AT20G survey at 20\,GHz (red) and the NVSS survey at 1.4\,GHz (blue) overlaid on an optical image of the field of PKS\,0048-12 (GLEAM J005026-120115). The 20\,GHz contours are elongated towards a bright, unresolved optical/IR object, WISE J0005025.57-120110.4, which has optical and mid-IR colours characteristic of a QSO. }
    \label{fig:pks0048-12}
\end{figure}

\noindent
{\bf GLEAM J005026-120115} (PKS\,0048-12) \\
A bright WISE source located 7\,arcsec from the 
radio centroid has mid-IR colours typical of QSO, and is also detected as a ROSAT X-ray source. Based on the radio morphology seen in the higher-resolution AT20G image, where the radio contours are extended in the direction of the optical object (see Figure \ref{fig:pks0048-12}), it appears probable that the strongly-scintillating low-frequency source is one hotspot of a radio-loud QSO. We therefore identify this WISE source as the likely host of the radio source PKS\,0048-12. \\

\noindent
{\bf GLEAM J005027-175237} (TXS\,0047-181) \\
There is a faint, uncatalogued WISE source (with W1 = $18.31\pm0.30$ mag, and undetected in the W2-W4 bands) offset by 6.1\, arcsec from the NVSS radio position. It is unclear at this stage whether this faint object is associated with the radio source. The WISE image shows other faint sources nearby, suggesting the possible presence of a distant cluster of galaxies. \\

\noindent
{\bf GLEAM J005039-102734 } (MRC\,0048-107) \\
A visual inspection of the WISE W1 image shows two mid-IR sources close to the FIRST radio position. The closest object listed in the WISE catalogue (WISE J005038.95-102739.4 with W1 = 15.78 mag) is 3.0 arcsec away, but a second uncatalogued object is closer to the FIRST radio centroid. Based on measurements of the WISE made by one of us (THJ), we adopt WISE values of W1 = $15.67\pm0.05$, W2 = $15.21\pm0.14$ and W3 =  $10.2\pm0.95$ mag.\ for the radio-source host (which is not detected in the W4 band). \\

\noindent
{\bf GLEAM J005536-095216} (MRC\,0053-101) \\
The host of this double source remains unidentified, though there is a faint uncatalogued WISE object close to the radio centroid that appears more prominent in W2 than in the W1 band. The closest catalogued WISE source (with W1 = $15.06\pm0.04$ mag) is 5.3\,arcsec away and is unlikely to be associated.  \\

\noindent
{\bf GLEAM J005734-012329} (3C 29) \\
This GLEAM source is associated with the nearby galaxy UGC\,595, and is resolved into a pair of double sources in FIRST. The radio position listed in columns 9 and 10 of Table \ref{tab:data2} is that of the 20\,GHz core source detected in the AT20G survey \citep{Murphy2010}, rather than the FIRST radio centroid. \\

\noindent
{\bf GLEAM J010159-105555} (TN\,J0102-07) \\
\cite{DeBreuck2002} obtained deep K-band imaging of the host of this ultra steep-spectrum radio source, and identified it with a faint (K $\sim19$\,mag) galaxy. 
There is no catalogued WISE source near the FIRST radio position, implying that the host is fainter than about 17.2 mag in the WISE W1 band (as expected from the \cite{DeBreuck2002} imaging result). \\

\noindent
{\bf GLEAM J011156-302623} (TXS\,0109-307) \\
We tentatively identify this radio source with a WISE object located 3.9\,arsec from the catalogued NVSS position. \\

\noindent
{\bf GLEAM J011406-100800 } (PMN\,J0114-1007) \\
A faint WISE source, WISE J011406.33-100802.6 with W1 = 16.80 mag, is located 3.7 arcsec from the FIRST radio position. Given the extended nature of the FIRST source (which makes the position of the radio centroid less certain), and the AGN-like colours of the WISE source (W1-W2 = 0.74 mag), we adopt the WISE object as the correct radio-source host. \\

\noindent
{\bf GLEAM J011645-041849 } (MRC\,0114-045) \\
An SDSS spectrum shows strong emission lines,including high-excitation [NeIII] and [Ne V] lines as well as broad Balmer emission lines. \\

\noindent
{\bf GLEAM J012227-042123 } (PKS 0119-04) \\
An SDSS spectrum is available for this radio-loud QSO (SDSS J012227.89-042127.1), with the SDSS redshift listed as z = 2.7838. This is inconsistent with the value of z = 1.925 published by \cite{Osmer1994} and is also inconsistent with the position of several of the emission lines seen in the SDSS spectrum. The listed SDSS redshift appears to be incorrect, and we therefore adopt the \cite{Osmer1994} redshift of z = 1.925 in this paper. \\

\noindent
{\bf GLEAM J012231-061953} (MRC\,0120-065) \\
We tentatively identify this triple FIRST source with a bright WISE objects located near the stronger (eastern) hotspot. Higher-resolution radio images (ideally at frequencies above 5\,GHz) would be useful to check this identification. \\

\noindent
{\bf GLEAM J012603-012356} (NGC\,547) \\
This GLEAM source corresponds to the southern hotspot of the nearby extended radio galaxy NGC\,547 \citep{odea1985}. The position listed in columns 9 and 10 of Table \ref{tab:data2} is that of the optical galaxy, rather than the FIRST radio centroid. \\

\noindent
{\bf GLEAM J013136-070354} (PKS\,0129-073) \\
The WISE W1 image shows a faint source near the radio centroid, but the closest catalogued WISE source (WISE J013136.64-070400.4 with W1 = 14.41 mag) is 4.2 arcsec away and appears unlikely to be a radio match. \\

\noindent
{\bf GLEAM J013919-124741} (TXS\,0136-130) \\
A faint uncatalogued WISE source is visible near the NVSS radio position. One of us (THJ) has measured the WISE magnitudes of this source as W1 = $18.29\pm0.32$\,mag and W2 = $17.26\pm0.69$\,mag. \\

\noindent
{\bf GLEAM J014237-074232 } (PKS\,0140-07) \\
There is another (apparently unrelated) radio source about 2 arcmin from the FIRST position, which may confuse the GLEAM image. No nearby counterpart is seen in the WISE W1 image, but a faint optical object (SDSS J014236.44-074303.5) may correspond to the host galaxy. \\

\noindent
{\bf GLEAM J015455-204023 } (MRC\,0152-209) \\
This z=1.92 radio galaxy (`the Dragonfly') is described by \cite{Emonts2015} as ``the most infrared-luminous high-redshift radio galaxy known in the southern hemisphere'', and is embedded within a large reservoir of molecular gas \citep{Emonts2016}. 

The host galaxy of this source is not listed in the WISE catalogue, probably because it is blended with a nearby star. An object close to the radio position is visible in the WISE image, and one of us (THJ) has measured the following deblended magnitudes for the WISE galaxy: 
W1 = 16.259$\pm$0.036 mag; 
W2 = 15.177$\pm$0.014 mag;     
W3 = 11.973$\pm$0.096 mag;  
W4 =  8.828$\pm$0.081 mag. 


\bsp	
\label{lastpage}
\end{document}